\documentclass[%
 reprint,
superscriptaddress,
 amsmath,amssymb,
 aps,
floatfix,
]{revtex4-2}

\usepackage{graphicx}
\usepackage{mathtools}
\usepackage{xcolor}
\usepackage{dcolumn}
\usepackage{bm}
\usepackage{hyperref}
\hypersetup{
colorlinks=true,
urlcolor= blue,
citecolor=blue,
linkcolor= blue,
}

\begin{document}

\title{Manipulating Random Lasing Correlations in Doped Liquid Crystals}


\author{Yiyang Zhi}
\email[]{yiyang\_zhi3@berkeley.edu}
\affiliation{Department of Physics, Case Western Reserve University, 2076 Adelbert Rd, Cleveland, Ohio 44106, USA}
\affiliation{Department of Electrical Engineering and Computer Sciences, University of California, Berkeley, California, USA}

\author{Andrew Lininger}
\author{Giuseppe Strangi}
\email[]{giuseppe.strangi@case.edu}
\affiliation{Department of Physics, Case Western Reserve University, 2076 Adelbert Rd, Cleveland, Ohio 44106, USA}
\affiliation{Department of Physics, NLHT Lab - University of Calabria and CNR-NANOTEC Istituto di Nanotecnologia, 87036-Rende, Italy}


\date{\today}

\begin{abstract}

Random lasers are highly configurable light sources that are promising for imaging and photonic integration. In this study, random lasing action was generated by optically pumping N-(4-Methoxybenzylidene)-4-butylaniline (MBBA) liquid crystals infiltrated with gold nanoparticles and laser dye (pyrromethene 597). By varying the pump energy near lasing threshold, we show that it is possible to control the intensity correlations between the random lasing modes. The correlations in the system were phenomenologically characterized using the Lévy statistics of the emission spectra survival function. We also find that correlations and persistence of lasing action are correlated. These results demonstrate the possibility to dynamically control a key physical feature of random lasers, which may find applications in biomedical settings and network communications.


\end{abstract}
\maketitle

\section{\label{sec: Introduction} Introduction}

Random lasers (RLs) have been the subject of intense study for the past few decades after Letokhov's pioneering paper in 1968 on generating light in an active medium filled with scatterers~\cite{cao_lasing_2003}. Traditional lasers rely on well-aligned resonant cavities (such as Fabry-Pérot) to generate coherence and produce laser action. On the other hand, in random lasing systems the lasing action stems from disordered scatterers and the resulting random fluctuations of dielectric constant ($\kappa$) in the spatial domain~\cite{perumbilavil_beaming_2018}. The subsequent amplification of photons via stimulated emission occurs as a result of the synergy between coherent amplification and multiple scattering in the medium~\cite{wiersma_physics_2008}. RLs act as a testbed for nonlinear and mesoscopic physics~\cite{merrill_fluctuations_2016}. They also could have promising applications in imaging~\cite{redding_speckle-free_2012}, cancer detection~\cite{polson_cancerous_2010}, and photonic integration~\cite{wiersma_smallest_2000}. 

RLs are intrinsically stochastic, generally exhibiting fluctuations in wavelengths, the number of output modes, and intensities due to strong modal interactions within the gain material~\cite{ignesti_experimental_2013, merrill_fluctuations_2016}. Over the past several years, researchers have investigated the correlations among output modes by analyzing their intensity distribution utilizing a phase-insensitive intensity feedback model. The probability density function (PDF) and, equivalently, the survival function of the lasing intensities have been found to obey a Lévy distribution --- power law decay --- near the lasing threshold at high (normalized) intensities~\cite{wu_statistics_2007, gomes_observation_2016, uppu_statistical_2010}. In the current context, the survival function means complementary cumulative distribution function. Its mathematical asymptotic form is given as
\begin{equation}
S(I) = 1- \int_{-\infty}^{I} f(I') \, \text{d}I' \sim \frac{1}{I^m}, \label{eq: functional}
\end{equation}
where $f(I')$ is the PDF of intensities and $S(I)$ denotes the value of survival function at intensity $I$; the decay exponent $m$ has been experimentally determined on the order of $1$ \cite{uppu_identification_2012, wu_statistical_2008, ignesti_experimental_2013}. 

Based on existing literature, the statistical regime of random lasing when pumped near threshold has been established as Lévy distribution~\cite{lepri_statistical_2007, ignesti_experimental_2013}. Nevertheless, to the authors' best knowledge, there has not been a detailed work on how the Lévy statistics (quantified by $m$) is modified when the pump energy is varied near threshold. Here we use a system with well-known random lasing action~\cite{ferjani_random_2008, ferjani_statistical_2008, perumbilavil_soliton-assisted_2016, lee_electrically_2015, wiersma_temperature-controlled_2002, ye_influence_2016}: nematic liquid crystals infiltrated by laser dye and gold nanoparticles. We experimentally show a general increase in the correlation of the random lasing modes for increasing pump energy near threshold by performing a survival analysis on the emitted modes. Our results further suggest that random lasing correlation and persistence of the modes are directly related. We provide preliminary evidence that one can control a key physical property of a RL near threshold, which could pave the way for improved imaging and photonic circuits technologies.

\begin{figure*}
\includegraphics{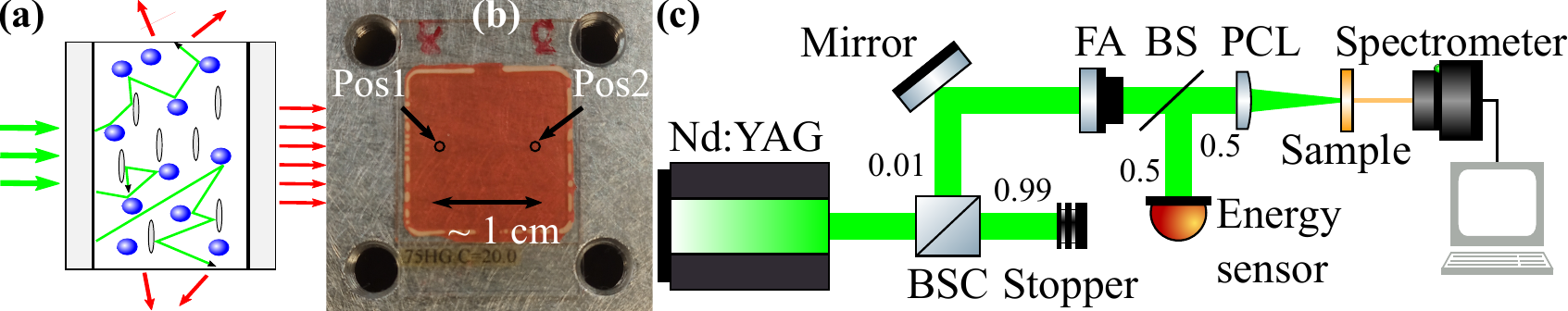}
\caption{\label{Fig: Set-up} \textbf{The random lasing system under consideration and relevant experimental set-up.} \textbf{(a)}, Schematic of random lasing occurring in doped LCs placed in a planar cell (gray rectangles). As the pump light (green arrows) traverses through the medium with PM597, it can scatter off of AuNPs (blue spheres) and LCs (gray ellipses). If a mode accumulates enough gain to surpass the loss, lasing occurs in all directions (red arrows). \textbf{(b)}, Experimental LC cell containing the prepared sample. The separation between Pos1 and Pos2 is more than $100$ times larger than the focused beam waist on the sample. The marked exposure points are drawn to scale. \textbf{(c)}, The output energy from the pulsed Nd:YAG laser is attenuated before being fine-tuned via a variable polarizer. The forward emission from the sample is analyzed by the spectrometer. The pump and spectrometer are synchronized to measure spectra from single pulses. BS(C), beam splitter (crystal); FA, fine attenuator; PCL, plano convex collimator.}
\end{figure*}
\section{\label{sec: Experimental set-up} System and Experimental set-up}


The RLs considered in this work were placed in a planar liquid crystal cell and optically excited by a pulsed laser to investigate its emission characteristics. Fig.~\ref{Fig: Set-up}\textbf{a} shows the schematic of the sample, which includes dispersed gold nanoparticles (AuNPs) and laser dye pyrromethene 597 (PM597) dissolved in N-(4-Methoxybenzylidene)-4-butylaniline (MBBA) liquid crystals (LCs). The LCs are in the nematic phase as the experiments were conducted at room temperature~\cite{rosta_ten_1987}. When the incident light diffuses through the medium, both LCs and AuNPs can act as Mie scatterers. Photons are subsequently ``trapped" in the cavities formed by those nanoscale scatterers. The trapping significantly increases the dwell time of the photons in the gain medium and the interaction with it. Lasing action subsequently occurs when modes attain enough gain to compensate for the loss.

AuNPs are often added to RLs to improve their performance~\cite{wan_pump-controlled_2019} since they support plasmonic enhancement resulting from subwavelength light confinement~\cite{wang_nanolasers_2017}. The localized surface plasmon resonance (LSPR) provided by dispersed AuNPs can collectively produce a resonant optical feedback for random lasing action because the emission spectrum from PM597 has significant overlap with the absorption peak of AuNPs~\cite{strangi_random_2006, bolis_nematicon-driven_2016}. Our sample was sonicated to colloidally disperse AuNPs and ensure homogeneous mixing. The \% by weight of AuNPs (diameter = $3 - 5$ nm) was chosen to be 0.2 \% to minimize quenching effects. See Supplementary Information S-I for the complete recipe of sample preparation and absoprtion spectra.


Fig.~\ref{Fig: Set-up}\textbf{b} is a top view of the prepared sample injected into the LC cell ($75 \, \mu \text{m}$ thickness, planar alignment, ITO coated and rubbed). The two black spots (Pos1 and Pos2) are drawn to indicate the exposure points to the pump laser, and the size of the dots are drawn to scale. The experimental data from these two regions serve to test the reproducibility of our results on the same system. The physical distance between Pos1 and Pos2 ($\sim 1$ cm) is much greater than the focused beam waist ($\simeq 70 \, \mu \text{m} $), rendering them independent from each other. In addition, they can be regarded as two different configurations of the same photonic system because the local spatial distributions of LCs and AuNPs can be different.

Fig.~\ref{Fig: Set-up}\textbf{c} is the schematic for the experimental set-up. To modulate the incident energy delivered, the light beam from a Q-switched Nd: YAG laser (Quantel laser, $\lambda = 532$ nm, $3$ ns pulse duration, and $f_{\text{rep.}} = 10$ Hz) was first attenuated to reach incident beam energies near lasing threshold. The energy of the linearly polarized pump beam is subsequently controlled by rotating a polarizer before illuminating on the sample. The forward emission from the sample is collected by a coupled optical fiber and fed into a spectrometer (Ocean Insights, HR4000CG-UV-NIR, wavelength resolution $= 0.27$ nm, integration time = $120$ ms). The pulse energy is measured by a pyroelectric energy sensor (Thorlabs, ES111C). 
The Q-switch of the pump laser and the spectrometer acquisition are synchronized such that one spectrum is taken for each pulse. The pump energy was tuned as follows. After determining the lasing threshold ($E_{\text{th.}}$) at Pos1, we increased the energy from $1.1 \, E_{\text{th.}}$ to $1.8 \, E_{\text{th.}}$ in step of $0.12\, E_{\text{th.}}$ (ramp up phase). At each step, $200$ single-shot emission spectra were collected, and we provided a 10-minute relaxation time before changing the pump energy. Staying at Pos1, we ramped back down to the original energy values in the same $0.12\, E_{\text{th.}}$ increment, recording spectra and leaving the sample idle in the same way as the ramp up phase. The entire procedure was repeated for Pos2.

\section{\label{sec: Results} Results and Discussion}

Survival analysis of the random lasing modes are performed at each pump energy values to extract the decay exponents ($m$), which reveal the Lévy statistics and correlations in the RL. We start the section by presenting the procedure for obtaining $m$. Then, we demonstrate the relationship between correlations and pump energy, cross-validating between the ramp up and down phases for two samples. Finally, we characterize the lasing persistence via the coefficient of variation (COV) and explore the synergy between correlation and persistence.


\subsection{\label{subsec: individual analysis} Analysis Techniques for a Single Energy Value}
\begin{figure*}
    \centering
    \includegraphics{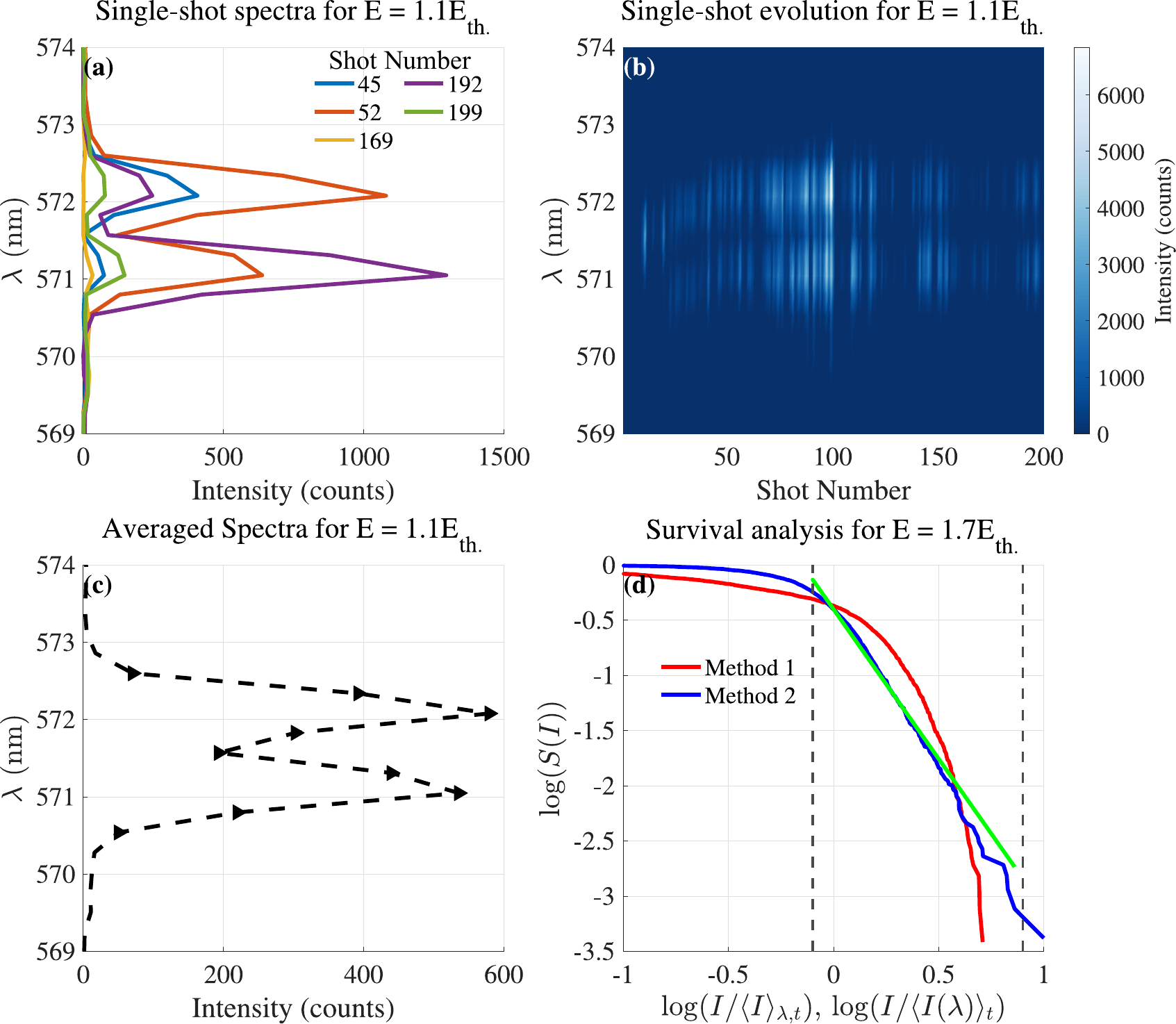}
    \caption{\textbf{Example survival analysis.} \textbf{(a)}, Five randomly sampled spectra of the sample emission, showing the huge fluctuations in intensity over the $20$ seconds acquisition time. The dominant modes at approximately $571$ and $572$ nm possess sub-nanometer width. These two features prove that we are observing random lasing in our system. \textbf{(b)}, The time-resolved spectra over $200$ shots. The two vague horizontal stripes reflect the two dominant modes in panel \textbf{a}. \textbf{(c)}, The temporal average of intensities at each wavelength of interest. The dark triangles mark the modes that pass the selection criteria and will be included in the survival analysis. \textbf{(d)}, The obtained survival function after appropriate normalization using two methods at incident energy $E = 1.7 E_{\text{th.}}$, shown here as a representative fit for all energy values. All fitted survival functions used in this study can be found in Supplementary Information S-III. A linear fit was performed in the region $[-0.1, 0.9]$, where we expect the \textit{fat tail} feature, to extract the correlation statistics between random lasing modes. The green line is the fit to method 2.}
    \label{Fig: example analysis}
\end{figure*}
We take the data from Pos1 at $E = 1.1 E_{\text{th.}} = 30.92 \, \mu$J to illustrate the analysis steps taken to obtain $m$ at a particular incident energy. The same procedure is repeated for all pump energies. Fig.~\ref{Fig: example analysis}\textbf{a} shows $5$ randomly sampled spectra out of all captured ones ($200$). Despite the large fluctuations in lasing intensities between spectra, two dominant, sub-nanometer output modes are observed at $571$ and $572$ nm. 

Fig.~\ref{Fig: example analysis}\textbf{b} provides an overview of the emission characteristics in the entire $200$ shots acquisition period. In the colormap, each vertical slice is a spectrum of the sample at that particular time instant. Note some slight red shifting (about $0.5$ nm) in the primary lasing modes during the first $50$ spectra. Emission wavelengths shifting is a signature property of RLs and is irrelevant to our subsequent analysis (see Supplementary Information S-II for other instances of emission wavelengths shifting). The emission quickly stabilizes to the two dominant modes that we are seeing in Fig.~\ref{Fig: example analysis}\textbf{a}. The persistent modes are manifested as light horizontal stripes in the single-shot spectra.

\begin{figure*}[t]
    \centering
    \includegraphics{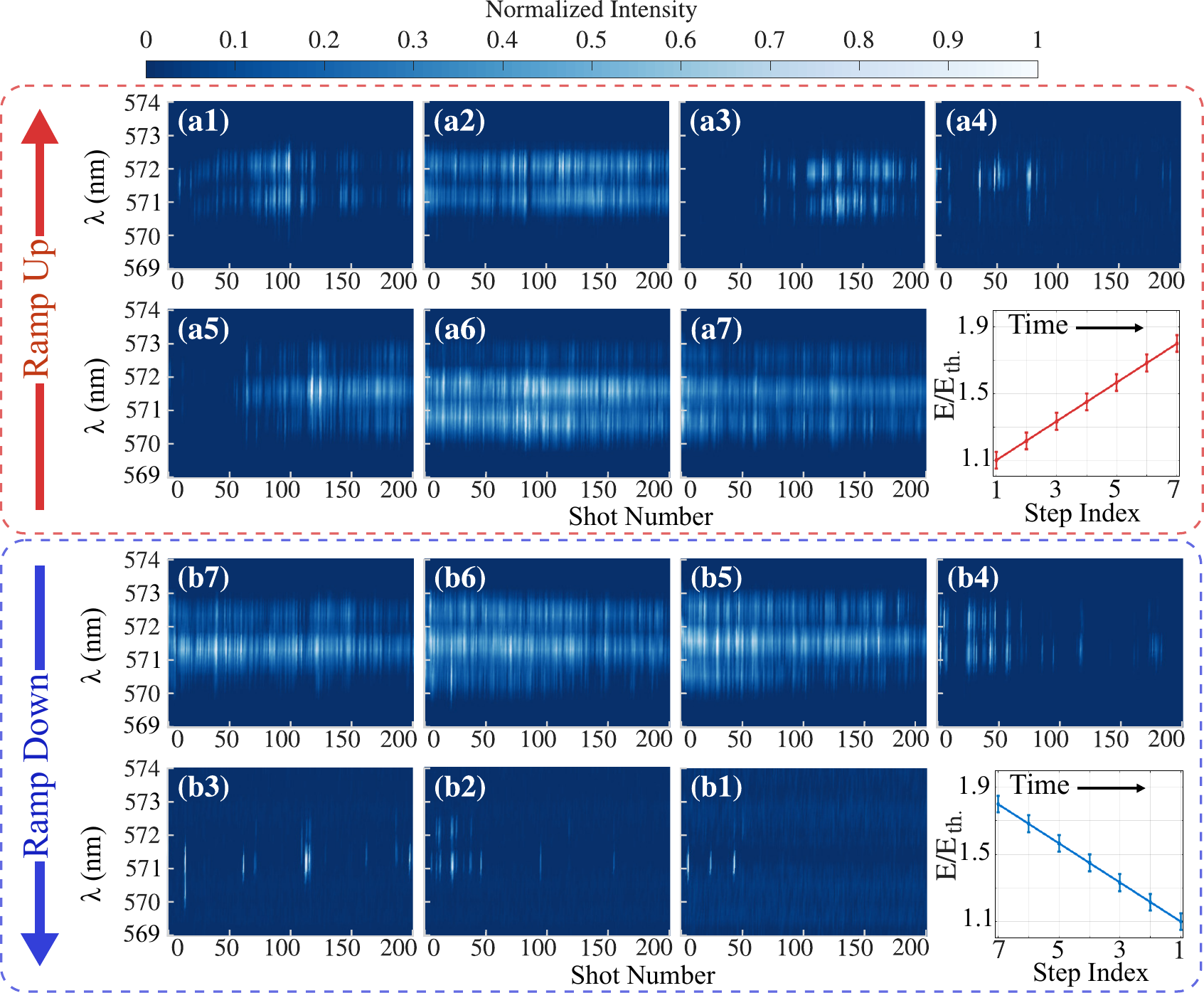}
    \caption{\textbf{The qualitative evolution of the RL for varying incident pump energy (Pos1).} \textbf{(a1) -- (a7),} Single-shot spectra in the ramp up phase. The energy range covered is $1.1 E_{\text{th.}} \rightarrow 1.8 E_{\text{th.}}$ in increment of $0.12 E_{\text{th.}}$. \textbf{(b7) -- (b1),} The corresponding spectra in the ramp down phase, starting at the highest energy. \textbf{(a8) and (b8),} Incident energy versus panel indices in the two phases, with the same index denotes the same energy values. The intensities are normalized with respect to each individual single-shot spectra. As the pump pulse energy reaches the maximum value and return to the original value, the lasing behavior is initially similar (compare \textbf{b7} -- \textbf{b5} with \textbf{a7} -- \textbf{a5}) but becomes markedly nonequivalent as we continue (\textbf{b3} -- \textbf{b1} versus \textbf{a3} -- \textbf{a1}). Thus, the RL exhibits hysteresis due to this gradual loss in lasing persistence.}
    \label{Fig: evolution pt1}
\end{figure*}

To select modes for constructing $S(I)$, we take the temporal average of the output intensity (Fig.~\ref{Fig: example analysis}\textbf{c}) over $200$ shots. Acceptable modes exhibit an average intensity $\geq 20$ counts (3$\sigma$ above the noise background). These modes are marked with dark triangles in Fig.~\ref{Fig: example analysis}\textbf{c}. Again, the temporal average reflects the long-standing modes at $571$ and $572$ nm. The chosen threshold value enables us to not only capture lasing modes when the lasing behavior is highly sporadic but also reject false peaks due to random noise.

In Fig.~\ref{Fig: example analysis}\textbf{d}, the selected intensities are pooled using two different methods following the work of Merrill, Cao, and Dufresne \cite{merrill_fluctuations_2016}. In method $1$ (M1), all the intensities are pooled before normalizing by the lump sum average, $\langle I \rangle_{t, \lambda}$, where $t$ labels the pulse counts. In method $2$ (M2), the intensities are first normalized at individual wavelength ($\langle I(\lambda) \rangle_{t}$) before pooling across wavelengths. In the case where the emission is completely independent across shots, the results generated from M1 and M2 will overlap.

We present the pooled survival function from $E = 1.7 E_{\text{th.}} = 48.90 \, \mu$J, since it possesses the representative linearity in the fitted interval for Lévy statistics. All of the fitted survival function used in the study can be found in Supplementary Information S-III. The curves resulting from M1 and M2 are in good agreement, and we chose the region $[-0.1, 0.9]$ in log-log space to use a linear fit to extract the Lévy statistics among selected lasing modes. The chosen region of moderately high normalized intensities is where we expect a asymptotic power law (Eq.~\ref{eq: functional}) \cite{merrill_fluctuations_2016, wu_statistics_2007, uppu_identification_2012}, with even higher intensities residing in the low fidelity region of $S(I)$. The fitting range remains identical for all energies, establishing a standard for the systematic comparison later.

The bright green line is a linear fit to the M2 line, with the decay exponent determined to be $m_{\text{fit}} = 2.71 \pm 0.02$ and $R^2 = 0.984$. The fit to the M1 line generates similar statistics. There exists some deviation between the model and data due to the finite sample size. Moreover, fitting a standard exponential decay curve of the form $ae^{bx} + c$ representing the free decay of photonic modes produces a curve that is identical to the bright green line on this scale, i.e., $b \ll 1$, showing that the power law relationship is the best description for the behavior of the system in this region on both physical and mathematical ground. As we show in Supplementary Information S-III, the $R^2$ values for all pump energy values remain above $0.85$, and exponential fits reduces to a linear one for the vast majority of data sets used in this study.

\subsection{\label{subsec: Hysteresis in the System} Hysteresis in the System}
\begin{figure*}
    \includegraphics{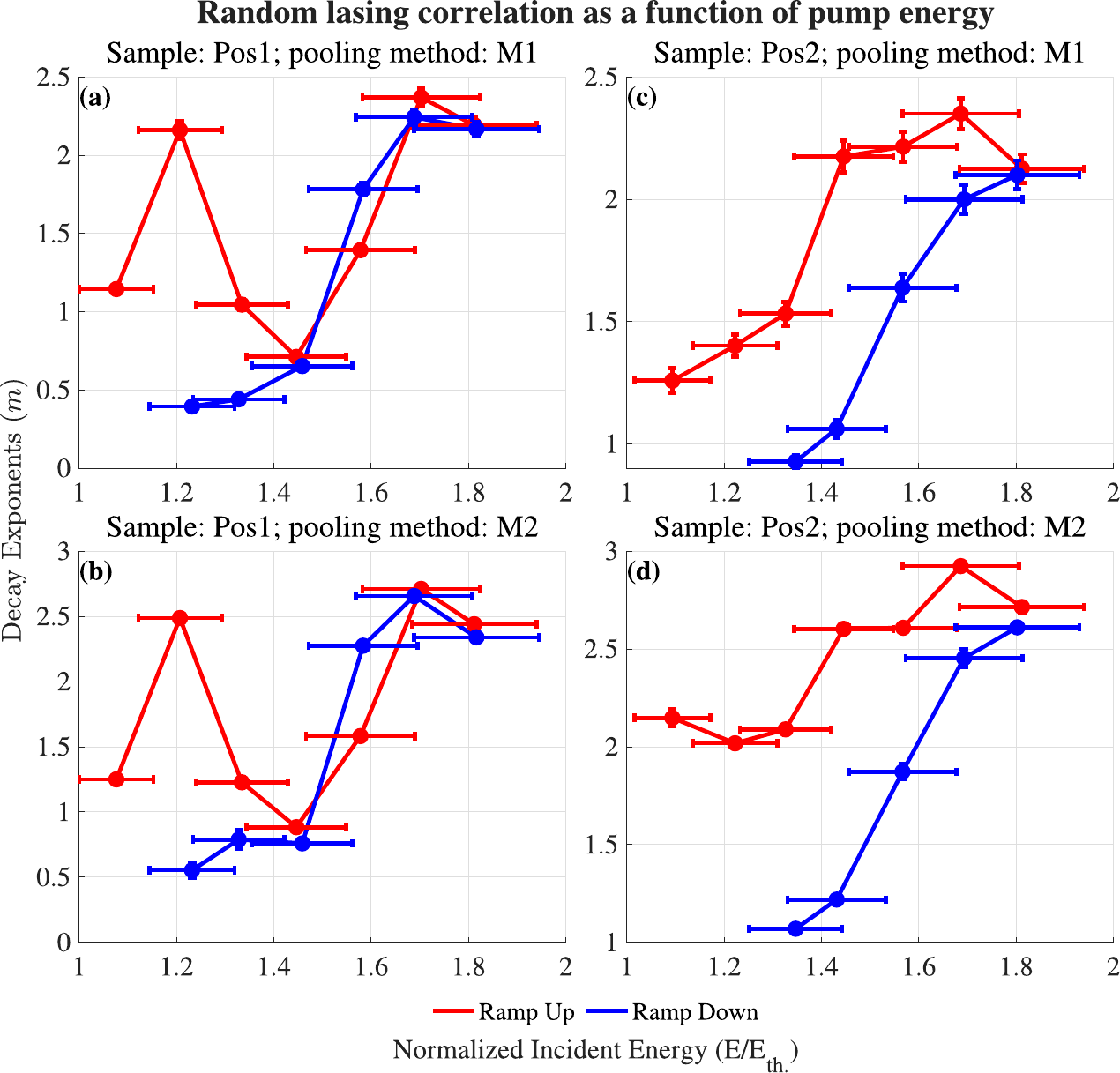}
    \caption{\textbf{Decay exponents as a function of incident energy compared across samples and methods.} \textbf{(a) and (b),} The response of decay exponents as we tune the pump pulse energy at Pos1 using M1 and M2 as our pooling method, respectively. \textbf{(c) and (d),} The same set of curves for Pos2. The trend of increasing in correlations following an increase in incident energy is preserved regardless of pooling methods or samples.}
    \label{Fig: correlations analysis}
\end{figure*}
Hysteresis can be observed in the RLs during sequential pump energy ramp up and down phases. Fig.~\ref{Fig: evolution pt1} records the evolution of lasing action for Pos1. Each panel is plotted in the same manner as Fig.~\ref{Fig: example analysis}\textbf{b}. Panels \textbf{a1} $\rightarrow$ \textbf{a7} are spectra collected at $1.1 E_{\text{th.}} \rightarrow 1.8 E_{\text{th.}}$ in the ramp up phase. Panels \textbf{b7} $\rightarrow$ \textbf{b1} show the spectra when the pump energy is $1.8 E_{\text{th.}} \rightarrow 1.1 E_{\text{th.}}$ in the ramp down phase. The same numeric indices in the two phases indicate the same pump energies within $5 \%$ difference. The pump energy at each index is shown in Fig.~\ref{Fig: evolution pt1}\textbf{a8} and \textbf{b8} for ramp up and down phase, respectively. The intensities are individually normalized for each set of spectra.

\begin{figure*}
    \includegraphics{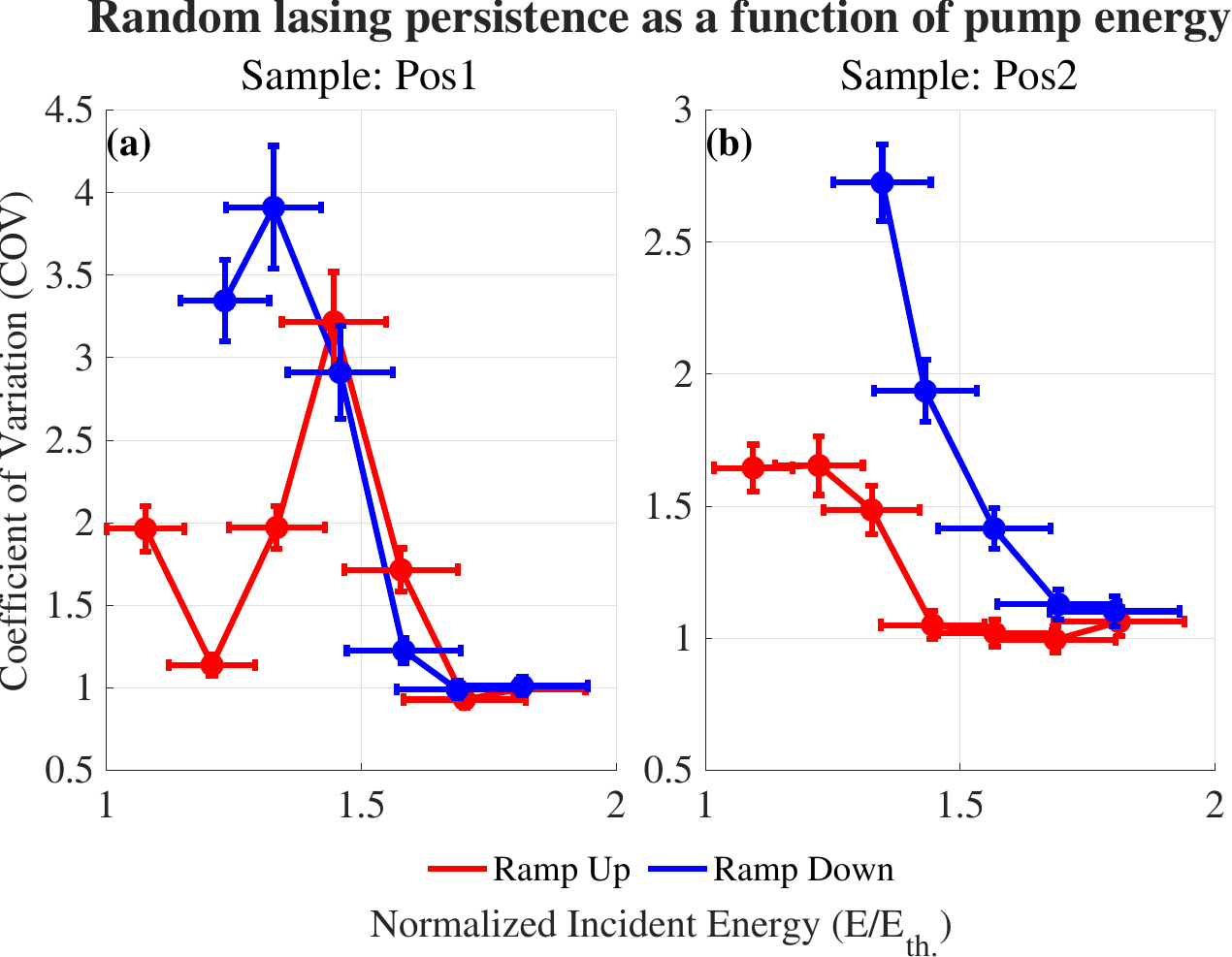}
    \caption{\textbf{Lasing persistence as a function of pump pulse energy.} \textbf{(a) and (b),} COV versus normalized incident energy for Pos1 and Pos2, respectively. We see a drop in COV for higher input intensity, which implies an increase in lasing persistence.}
    \label{Fig: interplay}
\end{figure*}

Progressing through the ramp phase (\textbf{a1} -- \textbf{a7}), we see that random lasing action, overall, becomes more persistent and possesses sharper contrast to the background with increasing incident energy. There is also an emergence of sidebands since there is more gain available in the system to simultaneously support competing modes. Ramping the energy down at the same position, the emission characteristics are very similar at higher pump energies (\textbf{b7} -- \textbf{b5}) but start to differentiate for lower energy values (\textbf{b3} -- \textbf{b1}). For the lowest energy values (past panel \textbf{b4}, $1.46 E_{\text{th.}}$), the lasing is highly sporadic.

Panel \textbf{a2} can be considered as an anomaly due the intense and consistent lasing at relatively low pump energy, breaking the general trend that we stated in the previous paragraph. The reason for this outlier is under active investigation, but one potential explanation would be the coincidental formation of a highly favorable local configuration of nanoscatterers for random lasing due to the energy and momentum transferred by the pump laser. As we will see in the next section, this outlier does not influence our overall analysis of intensity correlation as a function of pump energy. See Supplementary Information-II for more instances of such anomalous behavior.

Tracing the progression of the single-shot spectra, we observe a changing behavior when the sample is kept under continued excitation. The lasing action observed in the beginning of the cycle begins to cease until it completely disappears. One potential reason for this irreversibility could be the modification of local conditions by the pump laser when the system is excited for a prolonged period of time, even when significant relaxation time between each $200$ shot acquisition was allowed. 

We can exclude three causes of the observed hysteresis. First, it is unlikely to be caused by simply quenching or photo-bleaching of our sample, since the characteristic \textit{monotonic} decrease in intensity is not observed (see Supplementary Information-IV). Second, this cannot be due to a drift in pump laser energy over time, since the standard deviation for all indices is smaller than $2 \, \mu$J, or $\approx 0.07 E_{\text{th.}}$. Third, the red/blue shift in the emission spectra is not a necessary condition for hysteresis, as similar behavior is observed in additional samples without any wavelength shifting (see Supplementary Information S-V).

%
%
\subsection{\label{subsec: Correlations Analysis} Correlations Analysis}

The methods discussed in Sec. \ref{subsec: individual analysis} were employed at all incident energies to extract the decay exponents ($m$) and characterize the correlation of random lasing modes. Fig.~\ref{Fig: correlations analysis}\textbf{a} and~\ref{Fig: correlations analysis}\textbf{b} show $m$ versus normalized incident energy for Pos1 using M1 and M2, respectively; Fig.~\ref{Fig: correlations analysis}\textbf{c} and~\ref{Fig: correlations analysis}\textbf{d} are identical analysis done for Pos2. The lines associated with ramp up and down do not have equal number of points since for some low energy values in the ramp down phase, the resulting averaged mode intensities are too low to reliably construct the survival function, $S(I)$. Some vertical error bars are not visible at the shown scale.

All extracted exponents are on the order of $1$, corroborating existing literature. All four panels show that the random lasing correlations increase with increasing pump energy. This trend is consistent across both samples and pooling methods; more importantly, it is insensitive to the hysteretic behavior discussed in Sec.~\ref{subsec: Hysteresis in the System}. The anomalous lasing behavior at ($1.22 E_{\text{th.}}$) introduces a spike in correlations at this energy but does not influence the general trend. From $1.46E_{\text{th.}}$ to $1.8E_{\text{th.}}$ in both Pos1 and Pos2, the values extracted from the ramp up and down phases agree with each other well. The two methods of pooling intensities exhibit a similar trend, with minor discrepancy in the actual value of $m$. The increase in correlations is highly non-trivial because the exponents more than double when increasing the pump energy by merely $10 \, \mu$J, going from $1.46E_{\text{th.}}$ to $1.8E_{\text{th.}}$. This suggests that correlations shows high sensitivity to input energy near threshold.

\subsection{\label{sec: Interplay between correlations and persistence} Interplay between Correlations and Persistence}

The lasing persistence can be quantified by utilizing the coefficient of variation (COV):
\begin{equation}
\text{COV}(E) = \frac{\sigma(E)}{\mu(E)}\label{eq: COV definition},
\end{equation}
where $\mu(E)$ and $\sigma(E)$ are the mean and standard deviation of the lasing modes intensities over $200$ shots for a single pump energy $E$. Conceptually, COV is a normalized version of standard deviation, and larger COV implies more fluctuations exist in the underlying parent distribution.


Fig.~\ref{Fig: interplay}\textbf{a} and~\ref{Fig: interplay}\textbf{b} show the COV as a function of normalized incident energy for Pos1 and Pos2, respectively. Both samples exhibit similar COV values, which are generally $ > 1$, indicative of the large fluctuations in our RL intensity. The number of data points plotted for the four curves respects the fact that survival analysis was not performed for Pos1 at $1.1 E_{\text{th.}}$ and Pos2 at $1.1 E_{\text{th.}}$ and $1.22 E_{\text{th.}}$ in the ramp down phase.

Overall, we see that COV generally decreases with increasing pump energy. This can be qualitatively confirmed by the data in Fig.~\ref{Fig: evolution pt1} panels \textbf{a5} -- \textbf{a7} and \textbf{b5} -- \textbf{b7}, noting that the lasing is much more persistent at higher incident energy values. Similar to $m$, the COV shows good agreement between ramp up and down phases over the interval $1.46E_{\text{th.}}$ to $1.8E_{\text{th.}}$. Nevertheless, the measured persistence increases by at least $50 \%$ over the energy range covered for both samples from respective minimum. 

Moreover, comparing to Fig.~\ref{Fig: correlations analysis} reveals that relatively high values of COV concur with relatively low values of decay exponents near $1.46E_{\text{th.}}$; the opposite scenario occurs at around $1.8E_{\text{th.}}$. In panel a, the outlier at $1.22E_{\text{th.}}$ results in a drop in COV, further contributing to the argument that the correlations and persistence are positively correlated.


In Supplementary Information S-VI, we give a possible physical origin of the observed synergy between correlation and persistence by taking into account photon dwell time and mean transport length in the system \cite{uppu_persistent_2011, tiwari_frequency_2012}. Our existing evidence firmly demonstrated that the qualitative increase in correlations in accompanied by an increased persistence, evidenced by a significant decrease in the COV values.

\section{\label{sec: Conclusion} Conclusion}

In this study, we report on the possibility of controlling the correlation of random lasing modes by tuning the pump energy. This effect has been investigated in the proximity of the lasing threshold of a physical system consisted of gold nanoparticles and pyrromethene 597 laser dye dissolved in MBBA liquid crystals. Despite the notable hysteresis that arises in the system as we vary the incident pump energy, we have shown compelling evidence that there is an increase in output modes correlation with increasing pump energy. Furthermore, highly correlated modes show more persistent lasing action, classified by the coefficient of variation. 

For future directions, it will be of interest to experimentally control the persistent lasing interval by changing the diameter of the nanoparticles, which modulates the spectral overlap between the plasmon resonance(s) and the gain curve. To explore correlations from the quantum mechanical perspective --- intensity fluctuations, entanglement, and discord~\cite{starshynov_quantum_2016, lodahl_spatial_2005} ---, it would also be beneficial to measure single photon statistics ($g^{(2)}(\tau=0)$) and investigate the mutual coherence among lasing modes.

\section{\label{{sec: Acknowledgments}} Acknowledgments}

The authors acknowledge support from the Ohio Third Frontier Project “Research Cluster on Surfaces in Advanced Materials” at Case Western Reserve University.

\section{\label{{sec: Author Contributions}} Author Contributions}

G.S. and Y.Z. conceived the project. Y.Z. performed the measurements and data analysis. A.L. assisted with the experimental set-up and data collection. G.S. supervised the project. All authors contributed to the final manuscript.





\bibliography{reference.bib}

\providecommand{\noopsort}[1]{}\providecommand{\singleletter}[1]{#1}%
\begin{thebibliography}{29}%
\makeatletter
\providecommand \@ifxundefined [1]{%
 \@ifx{#1\undefined}
}%
\providecommand \@ifnum [1]{%
 \ifnum #1\expandafter \@firstoftwo
 \else \expandafter \@secondoftwo
 \fi
}%
\providecommand \@ifx [1]{%
 \ifx #1\expandafter \@firstoftwo
 \else \expandafter \@secondoftwo
 \fi
}%
\providecommand \natexlab [1]{#1}%
\providecommand \enquote  [1]{``#1''}%
\providecommand \bibnamefont  [1]{#1}%
\providecommand \bibfnamefont [1]{#1}%
\providecommand \citenamefont [1]{#1}%
\providecommand \href@noop [0]{\@secondoftwo}%
\providecommand \href [0]{\begingroup \@sanitize@url \@href}%
\providecommand \@href[1]{\@@startlink{#1}\@@href}%
\providecommand \@@href[1]{\endgroup#1\@@endlink}%
\providecommand \@sanitize@url [0]{\catcode `\\12\catcode `\$12\catcode
  `\&12\catcode `\#12\catcode `\^12\catcode `\_12\catcode `\%12\relax}%
\providecommand \@@startlink[1]{}%
\providecommand \@@endlink[0]{}%
\providecommand \url  [0]{\begingroup\@sanitize@url \@url }%
\providecommand \@url [1]{\endgroup\@href {#1}{\urlprefix }}%
\providecommand \urlprefix  [0]{URL }%
\providecommand \Eprint [0]{\href }%
\providecommand \doibase [0]{https://doi.org/}%
\providecommand \selectlanguage [0]{\@gobble}%
\providecommand \bibinfo  [0]{\@secondoftwo}%
\providecommand \bibfield  [0]{\@secondoftwo}%
\providecommand \translation [1]{[#1]}%
\providecommand \BibitemOpen [0]{}%
\providecommand \bibitemStop [0]{}%
\providecommand \bibitemNoStop [0]{.\EOS\space}%
\providecommand \EOS [0]{\spacefactor3000\relax}%
\providecommand \BibitemShut  [1]{\csname bibitem#1\endcsname}%
\let\auto@bib@innerbib\@empty
\bibitem [{\citenamefont {Cao}(2003)}]{cao_lasing_2003}%
  \BibitemOpen
  \bibfield  {author} {\bibinfo {author} {\bibfnamefont {H.}~\bibnamefont
  {Cao}},\ }\bibfield  {title} {\bibinfo {title} {Lasing in random media},\
  }\href {https://doi.org/10.1088/0959-7174/13/3/201} {\bibfield  {journal}
  {\bibinfo  {journal} {Waves in Random Media}\ }\textbf {\bibinfo {volume}
  {13}},\ \bibinfo {pages} {R1} (\bibinfo {year} {2003})}\BibitemShut {NoStop}%
\bibitem [{\citenamefont {Perumbilavil}\ \emph {et~al.}(2018)\citenamefont
  {Perumbilavil}, \citenamefont {Piccardi}, \citenamefont {Barboza},
  \citenamefont {Buchnev}, \citenamefont {Kauranen}, \citenamefont {Strangi},\
  and\ \citenamefont {Assanto}}]{perumbilavil_beaming_2018}%
  \BibitemOpen
  \bibfield  {author} {\bibinfo {author} {\bibfnamefont {S.}~\bibnamefont
  {Perumbilavil}}, \bibinfo {author} {\bibfnamefont {A.}~\bibnamefont
  {Piccardi}}, \bibinfo {author} {\bibfnamefont {R.}~\bibnamefont {Barboza}},
  \bibinfo {author} {\bibfnamefont {O.}~\bibnamefont {Buchnev}}, \bibinfo
  {author} {\bibfnamefont {M.}~\bibnamefont {Kauranen}}, \bibinfo {author}
  {\bibfnamefont {G.}~\bibnamefont {Strangi}},\ and\ \bibinfo {author}
  {\bibfnamefont {G.}~\bibnamefont {Assanto}},\ }\bibfield  {title} {\bibinfo
  {title} {Beaming random lasers with soliton control},\ }\href
  {https://doi.org/10.1038/s41467-018-06170-9} {\bibfield  {journal} {\bibinfo
  {journal} {Nature Communications}\ }\textbf {\bibinfo {volume} {9}},\
  \bibinfo {pages} {3863} (\bibinfo {year} {2018})}\BibitemShut {NoStop}%
\bibitem [{\citenamefont {Wiersma}(2008)}]{wiersma_physics_2008}%
  \BibitemOpen
  \bibfield  {author} {\bibinfo {author} {\bibfnamefont {D.~S.}\ \bibnamefont
  {Wiersma}},\ }\bibfield  {title} {\bibinfo {title} {The physics and
  applications of random lasers},\ }\href {https://doi.org/10.1038/nphys971}
  {\bibfield  {journal} {\bibinfo  {journal} {Nature Physics}\ }\textbf
  {\bibinfo {volume} {4}},\ \bibinfo {pages} {359} (\bibinfo {year}
  {2008})}\BibitemShut {NoStop}%
\bibitem [{\citenamefont {Merrill}\ \emph {et~al.}(2016)\citenamefont
  {Merrill}, \citenamefont {Cao},\ and\ \citenamefont
  {Dufresne}}]{merrill_fluctuations_2016}%
  \BibitemOpen
  \bibfield  {author} {\bibinfo {author} {\bibfnamefont {J.~W.}\ \bibnamefont
  {Merrill}}, \bibinfo {author} {\bibfnamefont {H.}~\bibnamefont {Cao}},\ and\
  \bibinfo {author} {\bibfnamefont {E.~R.}\ \bibnamefont {Dufresne}},\
  }\bibfield  {title} {\bibinfo {title} {Fluctuations and correlations of
  emission from random lasers},\ }\href
  {https://doi.org/10.1103/PhysRevA.93.021801} {\bibfield  {journal} {\bibinfo
  {journal} {Physical Review A}\ }\textbf {\bibinfo {volume} {93}},\ \bibinfo
  {pages} {021801} (\bibinfo {year} {2016})}\BibitemShut {NoStop}%
\bibitem [{\citenamefont {Redding}\ \emph {et~al.}(2012)\citenamefont
  {Redding}, \citenamefont {Choma},\ and\ \citenamefont
  {Cao}}]{redding_speckle-free_2012}%
  \BibitemOpen
  \bibfield  {author} {\bibinfo {author} {\bibfnamefont {B.}~\bibnamefont
  {Redding}}, \bibinfo {author} {\bibfnamefont {M.~A.}\ \bibnamefont {Choma}},\
  and\ \bibinfo {author} {\bibfnamefont {H.}~\bibnamefont {Cao}},\ }\bibfield
  {title} {\bibinfo {title} {Speckle-free laser imaging using random laser
  illumination},\ }\href {https://doi.org/10.1038/nphoton.2012.90} {\bibfield
  {journal} {\bibinfo  {journal} {Nature Photonics}\ }\textbf {\bibinfo
  {volume} {6}},\ \bibinfo {pages} {355} (\bibinfo {year} {2012})}\BibitemShut
  {NoStop}%
\bibitem [{\citenamefont {Polson}\ and\ \citenamefont
  {Vardeny}(2010)}]{polson_cancerous_2010}%
  \BibitemOpen
  \bibfield  {author} {\bibinfo {author} {\bibfnamefont {R.~C.}\ \bibnamefont
  {Polson}}\ and\ \bibinfo {author} {\bibfnamefont {Z.~V.}\ \bibnamefont
  {Vardeny}},\ }\bibfield  {title} {\bibinfo {title} {Cancerous tissue mapping
  from random lasing emission spectra},\ }\href
  {https://doi.org/10.1088/2040-8978/12/2/024010} {\bibfield  {journal}
  {\bibinfo  {journal} {Journal of Optics}\ }\textbf {\bibinfo {volume} {12}},\
  \bibinfo {pages} {024010} (\bibinfo {year} {2010})}\BibitemShut {NoStop}%
\bibitem [{\citenamefont {Wiersma}(2000)}]{wiersma_smallest_2000}%
  \BibitemOpen
  \bibfield  {author} {\bibinfo {author} {\bibfnamefont {D.}~\bibnamefont
  {Wiersma}},\ }\bibfield  {title} {\bibinfo {title} {The smallest random
  laser},\ }\href {https://doi.org/10.1038/35018184} {\bibfield  {journal}
  {\bibinfo  {journal} {Nature}\ }\textbf {\bibinfo {volume} {406}},\ \bibinfo
  {pages} {133} (\bibinfo {year} {2000})}\BibitemShut {NoStop}%
\bibitem [{\citenamefont {Ignesti}\ \emph {et~al.}(2013)\citenamefont
  {Ignesti}, \citenamefont {Tommasi}, \citenamefont {Fini}, \citenamefont
  {Lepri}, \citenamefont {Radhalakshmi}, \citenamefont {Wiersma},\ and\
  \citenamefont {Cavalieri}}]{ignesti_experimental_2013}%
  \BibitemOpen
  \bibfield  {author} {\bibinfo {author} {\bibfnamefont {E.}~\bibnamefont
  {Ignesti}}, \bibinfo {author} {\bibfnamefont {F.}~\bibnamefont {Tommasi}},
  \bibinfo {author} {\bibfnamefont {L.}~\bibnamefont {Fini}}, \bibinfo {author}
  {\bibfnamefont {S.}~\bibnamefont {Lepri}}, \bibinfo {author} {\bibfnamefont
  {V.}~\bibnamefont {Radhalakshmi}}, \bibinfo {author} {\bibfnamefont
  {D.}~\bibnamefont {Wiersma}},\ and\ \bibinfo {author} {\bibfnamefont
  {S.}~\bibnamefont {Cavalieri}},\ }\bibfield  {title} {\bibinfo {title}
  {Experimental and theoretical investigation of statistical regimes in random
  laser emission},\ }\href {https://doi.org/10.1103/PhysRevA.88.033820}
  {\bibfield  {journal} {\bibinfo  {journal} {Physical Review A}\ }\textbf
  {\bibinfo {volume} {88}},\ \bibinfo {pages} {033820} (\bibinfo {year}
  {2013})}\BibitemShut {NoStop}%
\bibitem [{\citenamefont {Wu}\ and\ \citenamefont
  {Cao}(2007)}]{wu_statistics_2007}%
  \BibitemOpen
  \bibfield  {author} {\bibinfo {author} {\bibfnamefont {X.}~\bibnamefont
  {Wu}}\ and\ \bibinfo {author} {\bibfnamefont {H.}~\bibnamefont {Cao}},\
  }\bibfield  {title} {\bibinfo {title} {Statistics of random lasing modes in
  weakly scattering systems},\ }\href {https://doi.org/10.1364/OL.32.003089}
  {\bibfield  {journal} {\bibinfo  {journal} {Optics Letters}\ }\textbf
  {\bibinfo {volume} {32}},\ \bibinfo {pages} {3089} (\bibinfo {year}
  {2007})}\BibitemShut {NoStop}%
\bibitem [{\citenamefont {Gomes}\ \emph {et~al.}(2016)\citenamefont {Gomes},
  \citenamefont {Raposo}, \citenamefont {Moura}, \citenamefont {Fewo},
  \citenamefont {Pincheira}, \citenamefont {Jerez}, \citenamefont {Maia},\ and\
  \citenamefont {de~Araújo}}]{gomes_observation_2016}%
  \BibitemOpen
  \bibfield  {author} {\bibinfo {author} {\bibfnamefont {A.~S.~L.}\
  \bibnamefont {Gomes}}, \bibinfo {author} {\bibfnamefont {E.~P.}\ \bibnamefont
  {Raposo}}, \bibinfo {author} {\bibfnamefont {A.~L.}\ \bibnamefont {Moura}},
  \bibinfo {author} {\bibfnamefont {S.~I.}\ \bibnamefont {Fewo}}, \bibinfo
  {author} {\bibfnamefont {P.~I.~R.}\ \bibnamefont {Pincheira}}, \bibinfo
  {author} {\bibfnamefont {V.}~\bibnamefont {Jerez}}, \bibinfo {author}
  {\bibfnamefont {L.~J.~Q.}\ \bibnamefont {Maia}},\ and\ \bibinfo {author}
  {\bibfnamefont {C.~B.}\ \bibnamefont {de~Araújo}},\ }\bibfield  {title}
  {\bibinfo {title} {Observation of {Lévy} distribution and replica symmetry
  breaking in random lasers from a single set of measurements},\ }\href
  {https://doi.org/10.1038/srep27987} {\bibfield  {journal} {\bibinfo
  {journal} {Scientific Reports}\ }\textbf {\bibinfo {volume} {6}},\ \bibinfo
  {pages} {27987} (\bibinfo {year} {2016})}\BibitemShut {NoStop}%
\bibitem [{\citenamefont {Uppu}\ and\ \citenamefont
  {Mujumdar}(2010)}]{uppu_statistical_2010}%
  \BibitemOpen
  \bibfield  {author} {\bibinfo {author} {\bibfnamefont {R.}~\bibnamefont
  {Uppu}}\ and\ \bibinfo {author} {\bibfnamefont {S.}~\bibnamefont
  {Mujumdar}},\ }\bibfield  {title} {\bibinfo {title} {Statistical fluctuations
  of coherent and incoherent intensity in random lasers with nonresonant
  feedback},\ }\href {https://doi.org/10.1364/OL.35.002831} {\bibfield
  {journal} {\bibinfo  {journal} {Optics Letters}\ }\textbf {\bibinfo {volume}
  {35}},\ \bibinfo {pages} {2831} (\bibinfo {year} {2010})}\BibitemShut
  {NoStop}%
\bibitem [{\citenamefont {Uppu}\ \emph {et~al.}(2012)\citenamefont {Uppu},
  \citenamefont {Tiwari},\ and\ \citenamefont
  {Mujumdar}}]{uppu_identification_2012}%
  \BibitemOpen
  \bibfield  {author} {\bibinfo {author} {\bibfnamefont {R.}~\bibnamefont
  {Uppu}}, \bibinfo {author} {\bibfnamefont {A.~K.}\ \bibnamefont {Tiwari}},\
  and\ \bibinfo {author} {\bibfnamefont {S.}~\bibnamefont {Mujumdar}},\
  }\bibfield  {title} {\bibinfo {title} {Identification of statistical regimes
  and crossovers in coherent random laser emission},\ }\href
  {https://doi.org/10.1364/OL.37.000662} {\bibfield  {journal} {\bibinfo
  {journal} {Optics Letters}\ }\textbf {\bibinfo {volume} {37}},\ \bibinfo
  {pages} {662} (\bibinfo {year} {2012})}\BibitemShut {NoStop}%
\bibitem [{\citenamefont {Wu}\ and\ \citenamefont
  {Cao}(2008)}]{wu_statistical_2008}%
  \BibitemOpen
  \bibfield  {author} {\bibinfo {author} {\bibfnamefont {X.}~\bibnamefont
  {Wu}}\ and\ \bibinfo {author} {\bibfnamefont {H.}~\bibnamefont {Cao}},\
  }\bibfield  {title} {\bibinfo {title} {Statistical studies of random-lasing
  modes and amplified spontaneous-emission spikes in weakly scattering
  systems},\ }\href {https://doi.org/10.1103/PhysRevA.77.013832} {\bibfield
  {journal} {\bibinfo  {journal} {Physical Review A}\ }\textbf {\bibinfo
  {volume} {77}},\ \bibinfo {pages} {013832} (\bibinfo {year}
  {2008})}\BibitemShut {NoStop}%
\bibitem [{\citenamefont {Lepri}\ \emph {et~al.}(2007)\citenamefont {Lepri},
  \citenamefont {Cavalieri}, \citenamefont {Oppo},\ and\ \citenamefont
  {Wiersma}}]{lepri_statistical_2007}%
  \BibitemOpen
  \bibfield  {author} {\bibinfo {author} {\bibfnamefont {S.}~\bibnamefont
  {Lepri}}, \bibinfo {author} {\bibfnamefont {S.}~\bibnamefont {Cavalieri}},
  \bibinfo {author} {\bibfnamefont {G.-L.}\ \bibnamefont {Oppo}},\ and\
  \bibinfo {author} {\bibfnamefont {D.~S.}\ \bibnamefont {Wiersma}},\
  }\bibfield  {title} {\bibinfo {title} {Statistical regimes of random laser
  fluctuations},\ }\href {https://doi.org/10.1103/PhysRevA.75.063820}
  {\bibfield  {journal} {\bibinfo  {journal} {Physical Review A}\ }\textbf
  {\bibinfo {volume} {75}},\ \bibinfo {pages} {063820} (\bibinfo {year}
  {2007})}\BibitemShut {NoStop}%
\bibitem [{\citenamefont {Ferjani}\ \emph
  {et~al.}(2008{\natexlab{a}})\citenamefont {Ferjani}, \citenamefont {Barna},
  \citenamefont {Luca}, \citenamefont {Versace},\ and\ \citenamefont
  {Strangi}}]{ferjani_random_2008}%
  \BibitemOpen
  \bibfield  {author} {\bibinfo {author} {\bibfnamefont {S.}~\bibnamefont
  {Ferjani}}, \bibinfo {author} {\bibfnamefont {V.}~\bibnamefont {Barna}},
  \bibinfo {author} {\bibfnamefont {A.~D.}\ \bibnamefont {Luca}}, \bibinfo
  {author} {\bibfnamefont {C.}~\bibnamefont {Versace}},\ and\ \bibinfo {author}
  {\bibfnamefont {G.}~\bibnamefont {Strangi}},\ }\bibfield  {title} {\bibinfo
  {title} {Random lasing in freely suspended dye-doped nematic liquid
  crystals},\ }\href {https://doi.org/10.1364/OL.33.000557} {\bibfield
  {journal} {\bibinfo  {journal} {Optics Letters}\ }\textbf {\bibinfo {volume}
  {33}},\ \bibinfo {pages} {557} (\bibinfo {year}
  {2008}{\natexlab{a}})}\BibitemShut {NoStop}%
\bibitem [{\citenamefont {Ferjani}\ \emph
  {et~al.}(2008{\natexlab{b}})\citenamefont {Ferjani}, \citenamefont
  {Sorriso-Valvo}, \citenamefont {De~Luca}, \citenamefont {Barna},
  \citenamefont {De~Marco},\ and\ \citenamefont
  {Strangi}}]{ferjani_statistical_2008}%
  \BibitemOpen
  \bibfield  {author} {\bibinfo {author} {\bibfnamefont {S.}~\bibnamefont
  {Ferjani}}, \bibinfo {author} {\bibfnamefont {L.}~\bibnamefont
  {Sorriso-Valvo}}, \bibinfo {author} {\bibfnamefont {A.}~\bibnamefont
  {De~Luca}}, \bibinfo {author} {\bibfnamefont {V.}~\bibnamefont {Barna}},
  \bibinfo {author} {\bibfnamefont {R.}~\bibnamefont {De~Marco}},\ and\
  \bibinfo {author} {\bibfnamefont {G.}~\bibnamefont {Strangi}},\ }\bibfield
  {title} {\bibinfo {title} {Statistical analysis of random lasing emission
  properties in nematic liquid crystals},\ }\href
  {https://doi.org/10.1103/PhysRevE.78.011707} {\bibfield  {journal} {\bibinfo
  {journal} {Physical Review E}\ }\textbf {\bibinfo {volume} {78}},\ \bibinfo
  {pages} {011707} (\bibinfo {year} {2008}{\natexlab{b}})}\BibitemShut
  {NoStop}%
\bibitem [{\citenamefont {Perumbilavil}\ \emph {et~al.}(2016)\citenamefont
  {Perumbilavil}, \citenamefont {Piccardi}, \citenamefont {Buchnev},
  \citenamefont {Kauranen}, \citenamefont {Strangi},\ and\ \citenamefont
  {Assanto}}]{perumbilavil_soliton-assisted_2016}%
  \BibitemOpen
  \bibfield  {author} {\bibinfo {author} {\bibfnamefont {S.}~\bibnamefont
  {Perumbilavil}}, \bibinfo {author} {\bibfnamefont {A.}~\bibnamefont
  {Piccardi}}, \bibinfo {author} {\bibfnamefont {O.}~\bibnamefont {Buchnev}},
  \bibinfo {author} {\bibfnamefont {M.}~\bibnamefont {Kauranen}}, \bibinfo
  {author} {\bibfnamefont {G.}~\bibnamefont {Strangi}},\ and\ \bibinfo {author}
  {\bibfnamefont {G.}~\bibnamefont {Assanto}},\ }\bibfield  {title} {\bibinfo
  {title} {Soliton-assisted random lasing in optically-pumped liquid
  crystals},\ }\href {https://doi.org/10.1063/1.4965852} {\bibfield  {journal}
  {\bibinfo  {journal} {Applied Physics Letters}\ }\textbf {\bibinfo {volume}
  {109}},\ \bibinfo {pages} {161105} (\bibinfo {year} {2016})}\BibitemShut
  {NoStop}%
\bibitem [{\citenamefont {Lee}\ \emph {et~al.}(2015)\citenamefont {Lee},
  \citenamefont {Lin}, \citenamefont {Guo}, \citenamefont {Lin}, \citenamefont
  {Lin}, \citenamefont {Zheng}, \citenamefont {Ma}, \citenamefont {Horng},
  \citenamefont {Sun},\ and\ \citenamefont {Huang}}]{lee_electrically_2015}%
  \BibitemOpen
  \bibfield  {author} {\bibinfo {author} {\bibfnamefont {C.-R.}\ \bibnamefont
  {Lee}}, \bibinfo {author} {\bibfnamefont {S.-H.}\ \bibnamefont {Lin}},
  \bibinfo {author} {\bibfnamefont {J.-W.}\ \bibnamefont {Guo}}, \bibinfo
  {author} {\bibfnamefont {J.-D.}\ \bibnamefont {Lin}}, \bibinfo {author}
  {\bibfnamefont {H.-L.}\ \bibnamefont {Lin}}, \bibinfo {author} {\bibfnamefont
  {Y.-C.}\ \bibnamefont {Zheng}}, \bibinfo {author} {\bibfnamefont {C.-L.}\
  \bibnamefont {Ma}}, \bibinfo {author} {\bibfnamefont {C.-T.}\ \bibnamefont
  {Horng}}, \bibinfo {author} {\bibfnamefont {H.-Y.}\ \bibnamefont {Sun}},\
  and\ \bibinfo {author} {\bibfnamefont {S.-Y.}\ \bibnamefont {Huang}},\
  }\bibfield  {title} {\bibinfo {title} {Electrically and thermally
  controllable nanoparticle random laser in a well-aligned dye-doped liquid
  crystal cell},\ }\href {https://doi.org/10.1364/OME.5.001469} {\bibfield
  {journal} {\bibinfo  {journal} {Optical Materials Express}\ }\textbf
  {\bibinfo {volume} {5}},\ \bibinfo {pages} {1469} (\bibinfo {year}
  {2015})}\BibitemShut {NoStop}%
\bibitem [{\citenamefont {Wiersma}\ and\ \citenamefont
  {Cavalieri}(2002)}]{wiersma_temperature-controlled_2002}%
  \BibitemOpen
  \bibfield  {author} {\bibinfo {author} {\bibfnamefont {D.~S.}\ \bibnamefont
  {Wiersma}}\ and\ \bibinfo {author} {\bibfnamefont {S.}~\bibnamefont
  {Cavalieri}},\ }\bibfield  {title} {\bibinfo {title} {Temperature-controlled
  random laser action in liquid crystal infiltrated systems},\ }\href
  {https://doi.org/10.1103/PhysRevE.66.056612} {\bibfield  {journal} {\bibinfo
  {journal} {Physical Review E}\ }\textbf {\bibinfo {volume} {66}},\ \bibinfo
  {pages} {056612} (\bibinfo {year} {2002})}\BibitemShut {NoStop}%
\bibitem [{\citenamefont {Ye}\ \emph {et~al.}(2016)\citenamefont {Ye},
  \citenamefont {Liu}, \citenamefont {Li}, \citenamefont {Feng}, \citenamefont
  {Cui},\ and\ \citenamefont {Lu}}]{ye_influence_2016}%
  \BibitemOpen
  \bibfield  {author} {\bibinfo {author} {\bibfnamefont {L.}~\bibnamefont
  {Ye}}, \bibinfo {author} {\bibfnamefont {B.}~\bibnamefont {Liu}}, \bibinfo
  {author} {\bibfnamefont {F.}~\bibnamefont {Li}}, \bibinfo {author}
  {\bibfnamefont {Y.}~\bibnamefont {Feng}}, \bibinfo {author} {\bibfnamefont
  {Y.}~\bibnamefont {Cui}},\ and\ \bibinfo {author} {\bibfnamefont
  {Y.}~\bibnamefont {Lu}},\ }\bibfield  {title} {\bibinfo {title} {The
  influence of {Ag} nanoparticles on random laser from dye-doped nematic liquid
  crystals},\ }\href {https://doi.org/10.1088/1612-2011/13/10/105001}
  {\bibfield  {journal} {\bibinfo  {journal} {Laser Physics Letters}\ }\textbf
  {\bibinfo {volume} {13}},\ \bibinfo {pages} {105001} (\bibinfo {year}
  {2016})}\BibitemShut {NoStop}%
\bibitem [{\citenamefont {Rosta}\ \emph {et~al.}(1987)\citenamefont {Rosta},
  \citenamefont {Kroó}, \citenamefont {Dolganov}, \citenamefont {Pacher},
  \citenamefont {Simkin}, \citenamefont {Török},\ and\ \citenamefont
  {Pépy}}]{rosta_ten_1987}%
  \BibitemOpen
  \bibfield  {author} {\bibinfo {author} {\bibfnamefont {L.}~\bibnamefont
  {Rosta}}, \bibinfo {author} {\bibfnamefont {N.}~\bibnamefont {Kroó}},
  \bibinfo {author} {\bibfnamefont {V.~K.}\ \bibnamefont {Dolganov}}, \bibinfo
  {author} {\bibfnamefont {P.}~\bibnamefont {Pacher}}, \bibinfo {author}
  {\bibfnamefont {V.~G.}\ \bibnamefont {Simkin}}, \bibinfo {author}
  {\bibfnamefont {G.}~\bibnamefont {Török}},\ and\ \bibinfo {author}
  {\bibfnamefont {G.}~\bibnamefont {Pépy}},\ }\bibfield  {title} {\bibinfo
  {title} {Ten {Phases} of {MBBA} ({A} {New} {Phase} {Diagram})},\ }\href
  {https://doi.org/10.1080/15421408708084223} {\bibfield  {journal} {\bibinfo
  {journal} {Molecular Crystals and Liquid Crystals}\ }\textbf {\bibinfo
  {volume} {144}},\ \bibinfo {pages} {297} (\bibinfo {year}
  {1987})}\BibitemShut {NoStop}%
\bibitem [{\citenamefont {Wan}\ and\ \citenamefont
  {Deng}(2019)}]{wan_pump-controlled_2019}%
  \BibitemOpen
  \bibfield  {author} {\bibinfo {author} {\bibfnamefont {Y.}~\bibnamefont
  {Wan}}\ and\ \bibinfo {author} {\bibfnamefont {L.}~\bibnamefont {Deng}},\
  }\bibfield  {title} {\bibinfo {title} {Pump-{Controlled} {Plasmonic} {Random}
  {Lasers} from {Dye}-{Doped} {Nematic} {Liquid} {Crystals} with {TiN}
  {Nanoparticles} in {Non}-{Oriented} {Cells}},\ }\href
  {https://doi.org/10.3390/app10010199} {\bibfield  {journal} {\bibinfo
  {journal} {Applied Sciences}\ }\textbf {\bibinfo {volume} {10}},\ \bibinfo
  {pages} {199} (\bibinfo {year} {2019})}\BibitemShut {NoStop}%
\bibitem [{\citenamefont {Wang}\ \emph {et~al.}(2017)\citenamefont {Wang},
  \citenamefont {Meng}, \citenamefont {Kildishev}, \citenamefont {Boltasseva},\
  and\ \citenamefont {Shalaev}}]{wang_nanolasers_2017}%
  \BibitemOpen
  \bibfield  {author} {\bibinfo {author} {\bibfnamefont {Z.}~\bibnamefont
  {Wang}}, \bibinfo {author} {\bibfnamefont {X.}~\bibnamefont {Meng}}, \bibinfo
  {author} {\bibfnamefont {A.~V.}\ \bibnamefont {Kildishev}}, \bibinfo {author}
  {\bibfnamefont {A.}~\bibnamefont {Boltasseva}},\ and\ \bibinfo {author}
  {\bibfnamefont {V.~M.}\ \bibnamefont {Shalaev}},\ }\bibfield  {title}
  {\bibinfo {title} {Nanolasers {Enabled} by {Metallic} {Nanoparticles}: {From}
  {Spasers} to {Random} {Lasers}},\ }\href
  {https://doi.org/10.1002/lpor.201700212} {\bibfield  {journal} {\bibinfo
  {journal} {Laser \& Photonics Reviews}\ }\textbf {\bibinfo {volume} {11}},\
  \bibinfo {pages} {1700212} (\bibinfo {year} {2017})}\BibitemShut {NoStop}%
\bibitem [{\citenamefont {Strangi}\ \emph {et~al.}(2006)\citenamefont
  {Strangi}, \citenamefont {Ferjani}, \citenamefont {Barna}, \citenamefont
  {Luca}, \citenamefont {Versace}, \citenamefont {Scaramuzza},\ and\
  \citenamefont {Bartolino}}]{strangi_random_2006}%
  \BibitemOpen
  \bibfield  {author} {\bibinfo {author} {\bibfnamefont {G.}~\bibnamefont
  {Strangi}}, \bibinfo {author} {\bibfnamefont {S.}~\bibnamefont {Ferjani}},
  \bibinfo {author} {\bibfnamefont {V.}~\bibnamefont {Barna}}, \bibinfo
  {author} {\bibfnamefont {A.~D.}\ \bibnamefont {Luca}}, \bibinfo {author}
  {\bibfnamefont {C.}~\bibnamefont {Versace}}, \bibinfo {author} {\bibfnamefont
  {N.}~\bibnamefont {Scaramuzza}},\ and\ \bibinfo {author} {\bibfnamefont
  {R.}~\bibnamefont {Bartolino}},\ }\bibfield  {title} {\bibinfo {title}
  {Random lasing and weak localization of light in dye-doped nematic liquid
  crystals},\ }\href {https://doi.org/10.1364/OE.14.007737} {\bibfield
  {journal} {\bibinfo  {journal} {Optics Express}\ }\textbf {\bibinfo {volume}
  {14}},\ \bibinfo {pages} {7737} (\bibinfo {year} {2006})}\BibitemShut
  {NoStop}%
\bibitem [{\citenamefont {Bolis}\ \emph {et~al.}(2016)\citenamefont {Bolis},
  \citenamefont {Virgili}, \citenamefont {Rajendran}, \citenamefont
  {Beeckman},\ and\ \citenamefont {Kockaert}}]{bolis_nematicon-driven_2016}%
  \BibitemOpen
  \bibfield  {author} {\bibinfo {author} {\bibfnamefont {S.}~\bibnamefont
  {Bolis}}, \bibinfo {author} {\bibfnamefont {T.}~\bibnamefont {Virgili}},
  \bibinfo {author} {\bibfnamefont {S.~K.}\ \bibnamefont {Rajendran}}, \bibinfo
  {author} {\bibfnamefont {J.}~\bibnamefont {Beeckman}},\ and\ \bibinfo
  {author} {\bibfnamefont {P.}~\bibnamefont {Kockaert}},\ }\bibfield  {title}
  {\bibinfo {title} {Nematicon-driven injection of amplified spontaneous
  emission into an optical fiber},\ }\href
  {https://doi.org/10.1364/OL.41.002245} {\bibfield  {journal} {\bibinfo
  {journal} {Optics Letters}\ }\textbf {\bibinfo {volume} {41}},\ \bibinfo
  {pages} {2245} (\bibinfo {year} {2016})}\BibitemShut {NoStop}%
\bibitem [{\citenamefont {Uppu}\ and\ \citenamefont
  {Mujumdar}(2011)}]{uppu_persistent_2011}%
  \BibitemOpen
  \bibfield  {author} {\bibinfo {author} {\bibfnamefont {R.}~\bibnamefont
  {Uppu}}\ and\ \bibinfo {author} {\bibfnamefont {S.}~\bibnamefont
  {Mujumdar}},\ }\bibfield  {title} {\bibinfo {title} {Persistent coherent
  random lasing using resonant scatterers},\ }\href
  {https://doi.org/10.1364/OE.19.023523} {\bibfield  {journal} {\bibinfo
  {journal} {Optics Express}\ }\textbf {\bibinfo {volume} {19}},\ \bibinfo
  {pages} {23523} (\bibinfo {year} {2011})}\BibitemShut {NoStop}%
\bibitem [{\citenamefont {Tiwari}\ \emph {et~al.}(2012)\citenamefont {Tiwari},
  \citenamefont {Uppu},\ and\ \citenamefont
  {Mujumdar}}]{tiwari_frequency_2012}%
  \BibitemOpen
  \bibfield  {author} {\bibinfo {author} {\bibfnamefont {A.~K.}\ \bibnamefont
  {Tiwari}}, \bibinfo {author} {\bibfnamefont {R.}~\bibnamefont {Uppu}},\ and\
  \bibinfo {author} {\bibfnamefont {S.}~\bibnamefont {Mujumdar}},\ }\bibfield
  {title} {\bibinfo {title} {Frequency behavior of coherent random lasing in
  diffusive resonant media},\ }\href
  {https://doi.org/10.1016/j.photonics.2012.04.001} {\bibfield  {journal}
  {\bibinfo  {journal} {Photonics and Nanostructures - Fundamentals and
  Applications}\ }\textbf {\bibinfo {volume} {10}},\ \bibinfo {pages} {416}
  (\bibinfo {year} {2012})}\BibitemShut {NoStop}%
\bibitem [{\citenamefont {Starshynov}\ \emph {et~al.}(2016)\citenamefont
  {Starshynov}, \citenamefont {Bertolotti},\ and\ \citenamefont
  {Anders}}]{starshynov_quantum_2016}%
  \BibitemOpen
  \bibfield  {author} {\bibinfo {author} {\bibfnamefont {I.}~\bibnamefont
  {Starshynov}}, \bibinfo {author} {\bibfnamefont {J.}~\bibnamefont
  {Bertolotti}},\ and\ \bibinfo {author} {\bibfnamefont {J.}~\bibnamefont
  {Anders}},\ }\bibfield  {title} {\bibinfo {title} {Quantum correlation of
  light scattered by disordered media},\ }\href
  {https://doi.org/10.1364/OE.24.004662} {\bibfield  {journal} {\bibinfo
  {journal} {Optics Express}\ }\textbf {\bibinfo {volume} {24}},\ \bibinfo
  {pages} {4662} (\bibinfo {year} {2016})}\BibitemShut {NoStop}%
\bibitem [{\citenamefont {Lodahl}\ \emph {et~al.}(2005)\citenamefont {Lodahl},
  \citenamefont {Mosk},\ and\ \citenamefont {Lagendijk}}]{lodahl_spatial_2005}%
  \BibitemOpen
  \bibfield  {author} {\bibinfo {author} {\bibfnamefont {P.}~\bibnamefont
  {Lodahl}}, \bibinfo {author} {\bibfnamefont {A.~P.}\ \bibnamefont {Mosk}},\
  and\ \bibinfo {author} {\bibfnamefont {A.}~\bibnamefont {Lagendijk}},\
  }\bibfield  {title} {\bibinfo {title} {Spatial {Quantum} {Correlations} in
  {Multiple} {Scattered} {Light}},\ }\href
  {https://doi.org/10.1103/PhysRevLett.95.173901} {\bibfield  {journal}
  {\bibinfo  {journal} {Physical Review Letters}\ }\textbf {\bibinfo {volume}
  {95}},\ \bibinfo {pages} {173901} (\bibinfo {year} {2005})}\BibitemShut
  {NoStop}%
\end{thebibliography}%


\providecommand{\noopsort}[1]{}\providecommand{\singleletter}[1]{#1}%
\begin{thebibliography}{3}%
\makeatletter
\providecommand \@ifxundefined [1]{%
 \@ifx{#1\undefined}
}%
\providecommand \@ifnum [1]{%
 \ifnum #1\expandafter \@firstoftwo
 \else \expandafter \@secondoftwo
 \fi
}%
\providecommand \@ifx [1]{%
 \ifx #1\expandafter \@firstoftwo
 \else \expandafter \@secondoftwo
 \fi
}%
\providecommand \natexlab [1]{#1}%
\providecommand \enquote  [1]{``#1''}%
\providecommand \bibnamefont  [1]{#1}%
\providecommand \bibfnamefont [1]{#1}%
\providecommand \citenamefont [1]{#1}%
\providecommand \href@noop [0]{\@secondoftwo}%
\providecommand \href [0]{\begingroup \@sanitize@url \@href}%
\providecommand \@href[1]{\@@startlink{#1}\@@href}%
\providecommand \@@href[1]{\endgroup#1\@@endlink}%
\providecommand \@sanitize@url [0]{\catcode `\\12\catcode `\$12\catcode
  `\&12\catcode `\#12\catcode `\^12\catcode `\_12\catcode `\%12\relax}%
\providecommand \@@startlink[1]{}%
\providecommand \@@endlink[0]{}%
\providecommand \url  [0]{\begingroup\@sanitize@url \@url }%
\providecommand \@url [1]{\endgroup\@href {#1}{\urlprefix }}%
\providecommand \urlprefix  [0]{URL }%
\providecommand \Eprint [0]{\href }%
\providecommand \doibase [0]{https://doi.org/}%
\providecommand \selectlanguage [0]{\@gobble}%
\providecommand \bibinfo  [0]{\@secondoftwo}%
\providecommand \bibfield  [0]{\@secondoftwo}%
\providecommand \translation [1]{[#1]}%
\providecommand \BibitemOpen [0]{}%
\providecommand \bibitemStop [0]{}%
\providecommand \bibitemNoStop [0]{.\EOS\space}%
\providecommand \EOS [0]{\spacefactor3000\relax}%
\providecommand \BibitemShut  [1]{\csname bibitem#1\endcsname}%
\let\auto@bib@innerbib\@empty
\bibitem [{\citenamefont {Demchenko}(2020)}]{demchenko_photobleaching_2020}%
  \BibitemOpen
  \bibfield  {author} {\bibinfo {author} {\bibfnamefont {A.~P.}\ \bibnamefont
  {Demchenko}},\ }\bibfield  {title} {\bibinfo {title} {Photobleaching of
  organic fluorophores: quantitative characterization, mechanisms,
  protection},\ }\href {https://doi.org/10.1088/2050-6120/ab7365} {\bibfield
  {journal} {\bibinfo  {journal} {Methods and Applications in Fluorescence}\
  }\textbf {\bibinfo {volume} {8}},\ \bibinfo {pages} {022001} (\bibinfo {year}
  {2020})}\BibitemShut {NoStop}%
\bibitem [{\citenamefont {Tiwari}\ \emph {et~al.}(2012)\citenamefont {Tiwari},
  \citenamefont {Uppu},\ and\ \citenamefont
  {Mujumdar}}]{tiwari_frequency_2012}%
  \BibitemOpen
  \bibfield  {author} {\bibinfo {author} {\bibfnamefont {A.~K.}\ \bibnamefont
  {Tiwari}}, \bibinfo {author} {\bibfnamefont {R.}~\bibnamefont {Uppu}},\ and\
  \bibinfo {author} {\bibfnamefont {S.}~\bibnamefont {Mujumdar}},\ }\bibfield
  {title} {\bibinfo {title} {Frequency behavior of coherent random lasing in
  diffusive resonant media},\ }\href
  {https://doi.org/10.1016/j.photonics.2012.04.001} {\bibfield  {journal}
  {\bibinfo  {journal} {Photonics and Nanostructures - Fundamentals and
  Applications}\ }\textbf {\bibinfo {volume} {10}},\ \bibinfo {pages} {416}
  (\bibinfo {year} {2012})}\BibitemShut {NoStop}%
\bibitem [{\citenamefont {Uppu}\ \emph {et~al.}(2012)\citenamefont {Uppu},
  \citenamefont {Tiwari},\ and\ \citenamefont
  {Mujumdar}}]{uppu_identification_2012}%
  \BibitemOpen
  \bibfield  {author} {\bibinfo {author} {\bibfnamefont {R.}~\bibnamefont
  {Uppu}}, \bibinfo {author} {\bibfnamefont {A.~K.}\ \bibnamefont {Tiwari}},\
  and\ \bibinfo {author} {\bibfnamefont {S.}~\bibnamefont {Mujumdar}},\
  }\bibfield  {title} {\bibinfo {title} {Identification of statistical regimes
  and crossovers in coherent random laser emission},\ }\href
  {https://doi.org/10.1364/OL.37.000662} {\bibfield  {journal} {\bibinfo
  {journal} {Optics Letters}\ }\textbf {\bibinfo {volume} {37}},\ \bibinfo
  {pages} {662} (\bibinfo {year} {2012})}\BibitemShut {NoStop}%
\end{thebibliography}%

\end{document}


\preprint{APS/123-QED}

\title{Supplementary Information to: Manipulating Random Lasing Correlations in Doped Liquid Crystals}

\author{Yiyang Zhi}
\email[]{yiyang\_zhi3@berkeley.edu}
\affiliation{Department of Physics, Case Western Reserve University, 2076 Adelbert Rd, Cleveland, Ohio 44106, USA}
\affiliation{Department of Electrical Engineering and Computer Sciences, University of California, Berkeley, California, USA}

\author{Andrew Lininger}
\author{Giuseppe Strangi}
\email[]{giuseppe.strangi@case.edu}
\affiliation{Department of Physics, Case Western Reserve University, 2076 Adelbert Rd, Cleveland, Ohio 44106, USA}
\affiliation{Department of Physics, NLHT Lab - University of Calabria and CNR-NANOTEC Istituto di Nanotecnologia, 87036-Rende, Italy}

\date{\today}

\maketitle

\section{Sample preparation and characterization}

\begin{figure}
    \centering
    \includegraphics{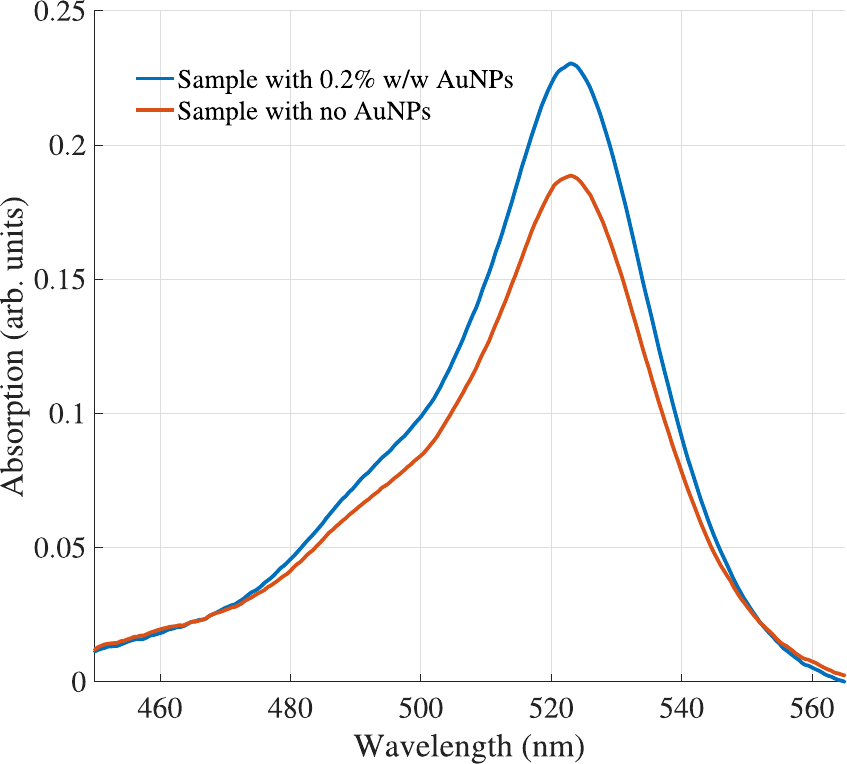}
    \caption{\textbf{UV-visible spectroscopy of the prepared RLs.} The blue curve is the absorption of the sample with AuNPs added (used in experiments), and the orange curve is the control with no AuNPs added but is otherwise the same. The sample exhibits absorption peak at $\sim 523$ nm that is $21.7 \%$ higher than the control, demonstrating that AuNPs are well dispersed in the solution. }
    \label{Fig: UV-Vis absorption}
\end{figure}

The MBBA LCs (Sigma-Aldrich) used for fabricating our random lasers were first transferred to vials using a pipette. Then, 0.2 \% by weight (w/w) of PM 597 (Exciton, Luxottica) was added to the LCs and sonicated for $10$ minutes. A pipette was used to add 0.2 \% w/w AuNPs (Sigma-Aldrich, dodecanethiol functionalized in toluene, diameter = $3$--$5$ nm, monodispersive) to the LCs/PM 597 mixture. The excess toluene was removed by heating the sample to $100$ $^\circ \text{C}$ for $10$ minutes interval and monitoring the mass of the sample until it stopped changing. Before performing experiments, the samples were sonicated for $\sim 30$ minutes. Ultraviolet–visible spectroscopy was performed on the samples to ensure that the AuNPs are well dispersed in the system. Fig.~\ref{Fig: UV-Vis absorption} shows the absorption data for the samples used in experiments (blue curve) and control (orange curve) solution that is identical to the samples except the addition of AuNPs. The absorption peak for the sample with AuNPs are $21.7 \%$ higher than that of control, which provides explicit evidence that the AuNPs are well dispersed in the LCs/PM 597 mixture. Before and after experiments, the samples were examined under polarized light microscopy and did not exhibit visible changes, which rules out the possibility of LCs cells damaged by the laser.  



\section{Random time series spectra}

\begin{figure*}
    \centering
    \includegraphics{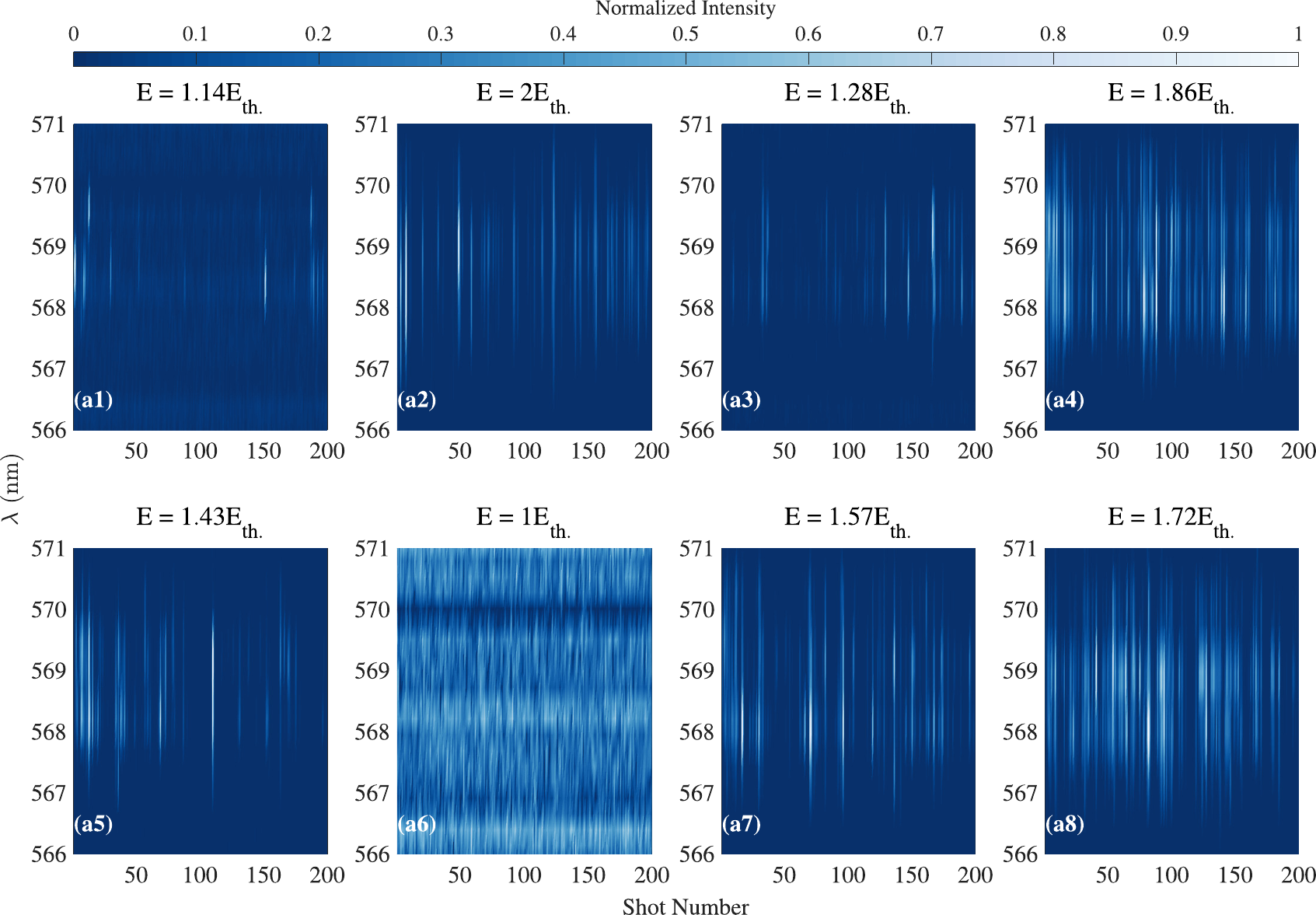}
    \caption{\textbf{The emission evolution of the RLs for permutated energy values.} \textbf{(a1) -- (a8),} Single-shot spectra of the system from $\sim E_{\text{th.}}$ to $2E_{\text{th.}}$. Increasing indices mean later exposure times. Randomized pump energies are shown in the titles for each panel. There exists emission wavelengths shifting for \textbf{a2} and \textbf{a5}, and panel \textbf{a8} can be treated as an anomaly since it exhibits more intense lasing compared to those spectra taken at higher energies. The wait time between adjacent acquisition is $10$ minutes.}
    \label{Fig: evolution_permutated}
\end{figure*}

We prepared an additional sample in identical ways compared to the that used in the main text and pump the sample from around $\sim E_{\text{th.}}$ to $2E_{\text{th.}}$ in the step of $0.14E_{\text{th.}}$. Here, we randomly permutated the order of the energies that was incident on the RLs. From this set of data, we aim to show that the coincidental formations of nano-scatterers configurations and emission wavelengths shifting are reproducible when we vary the pump energies differently. Fig.~\ref{Fig: evolution_permutated} is the evolution of the RLs single-shot spectra. The panels are ordered in the temporal exposure order, i.e., panel \textbf{a2} was acquired after \textbf{a1}. The titles of each panel show the corresponding pump energies.


In the spectra evolution presented in the main text, we noted an anomalous behavior where intense lasing behaviors are observed at relatively low pump energy. Here, Fig.~\ref{Fig: evolution_permutated}\textbf{a8} ($E=1.72E_{\text{{th.}}}$) can be considered as another anomaly, albeit to a lesser extent, since it exhibits more frequent lasing bursts compared to Fig.~\ref{Fig: evolution_permutated}\textbf{a2} ($E=2E_{\text{{th.}}}$). In addition, there is visible redshift in emission wavelengths in Fig.~\ref{Fig: evolution_permutated}\textbf{a2} within the first $50$ shots of acquisition. Therefore, we provide further evidence that these spectral features are owing to the stochastic nature of the RLs. As discussed in the main text, they do not have an effect on the general trend that we have established for correlations and coefficient of variation.

\section{Justification for the Lévy distribution fit}

Here we show that a linear relationship adequately describes the calculated survival function ($S(I)$) at large, normalized emission intensities ($\langle I \rangle_{t, \lambda}$ and $\langle I(\lambda) \rangle_{t}$) in log-log scale. As mentioned in the main text, we utilized two methods to combine the output modes. In method $1$ (M1), all the intensities are pooled and normalized by the lump sum average. In method $2$ (M2), the intensities are first normalized at individual wavelength before pooling.

\begin{figure}
    \centering
    \includegraphics{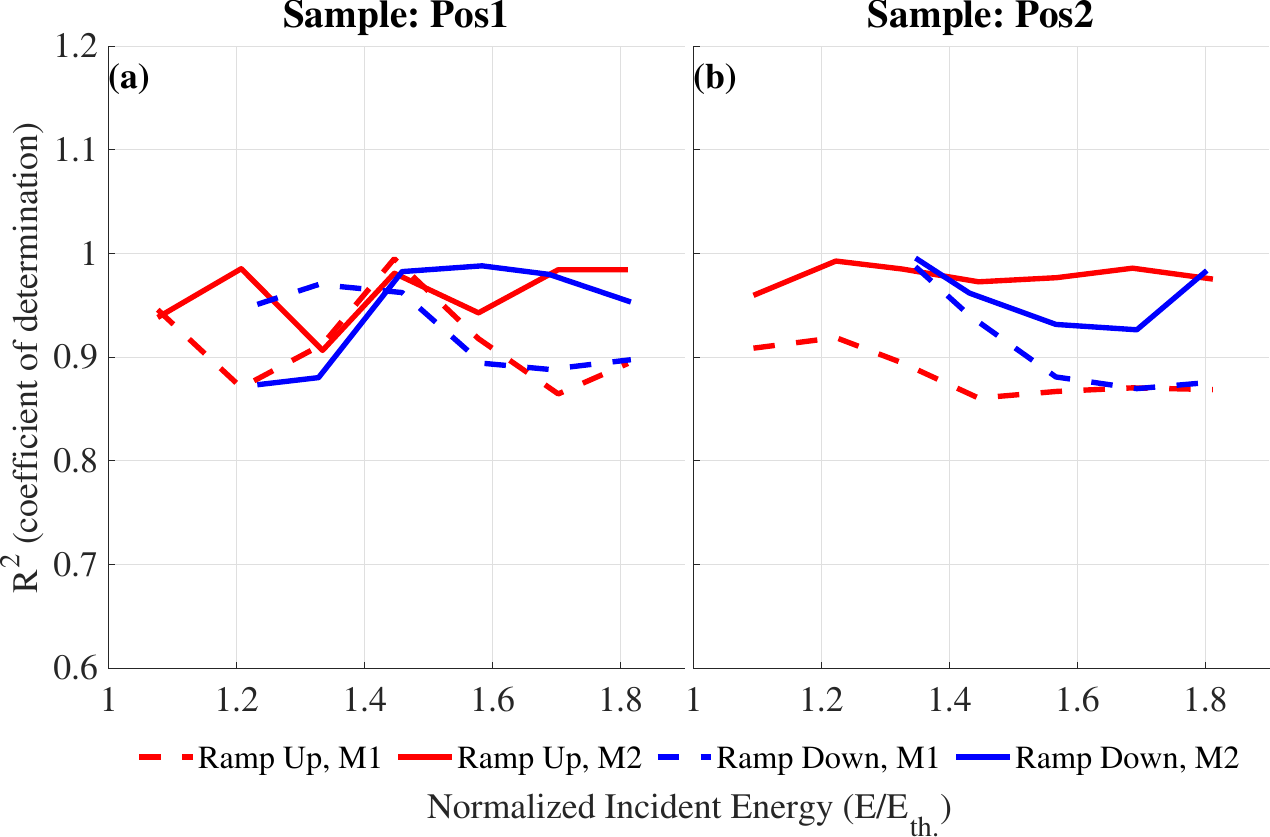}
    \caption{\textbf{Goodness of fit for the linear model used to extract the decay exponents.} \textbf{(a) and (b),} $R^2$ (coefficient of determination) as a function of normalized energy for Pos1 and Pos2, respectively. Regardless of which pooling method or energy varying direction is used, $R^2$ remains at least $0.85$. This provides reasonable justification for using Lévy statistics to describe the intensity correlation of our RLs.}
    \label{Fig: adjustedR_sqr}
\end{figure}

Fig. \ref{Fig: adjustedR_sqr} shows the $R^2$ of the linear regression performed on all sets of $S(I)$ versus normalized intensities over the interval $[-0.1, 0.9]$ in log-log space. The horizontal error bar is omitted here for clarity. Within each sample, $R^2$ values stay roughly constant with respect to energy, and there is good agreement between two pooling methods and energy varying directions. All fits produced a $R^2$ value of least $0.85$. Thus, a linear model reasonably describes the the behavior of $S(I)$ in the aforementioned interval, and we capturing the \textit{fat tail} of the Lévy distribution in our survival analysis of the single-shot spectra.

\begin{figure*}
    \centering
    \includegraphics{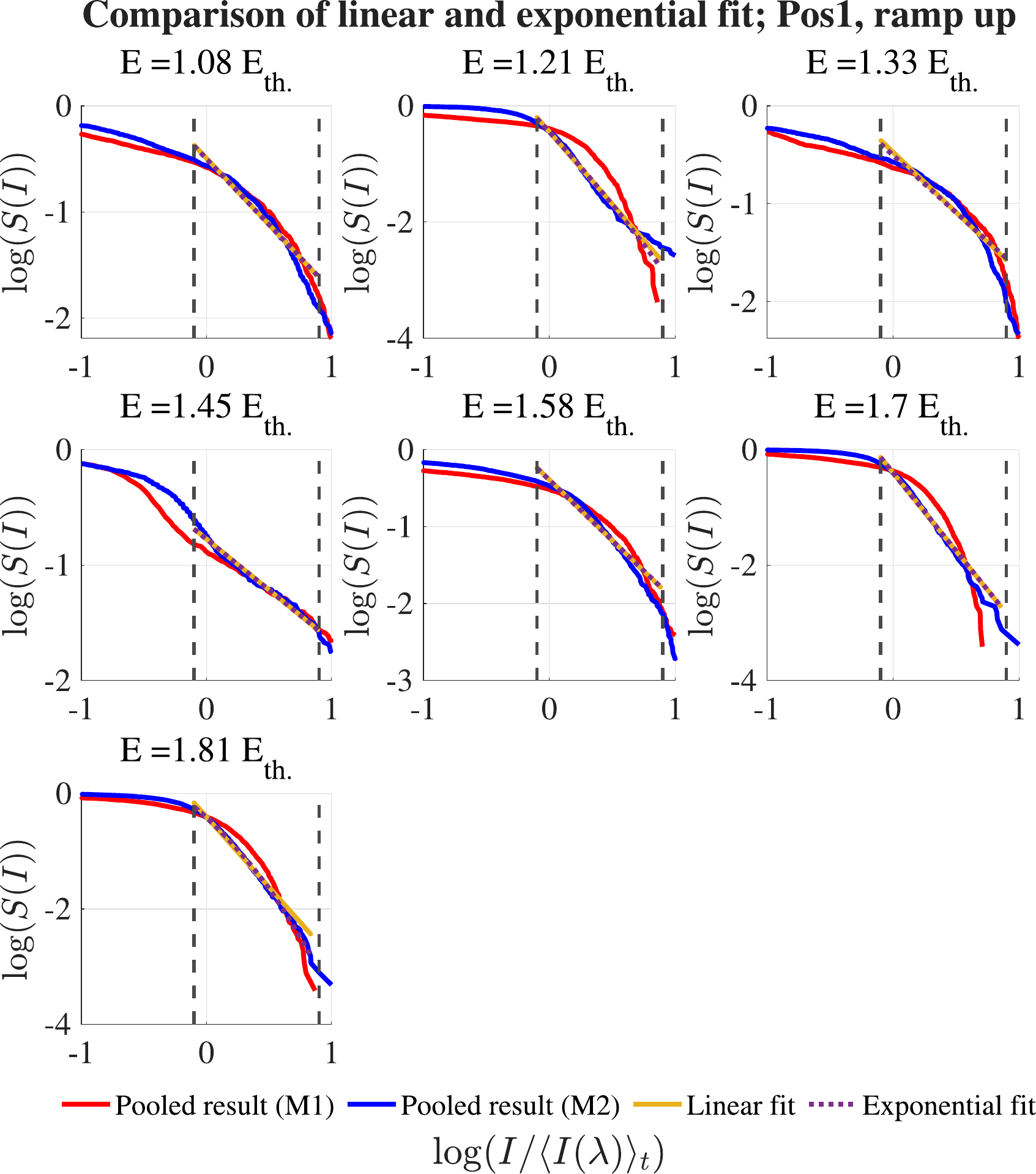}
    \caption{\textbf{Comparison of linear and exponential fit for Pos1, ramp up.} The fit is performed for M2. At the scale shown, linear fit (yellow) and exponential fit (purple) are identical except the last energy, $E = 1.81E_{\text{th.}}$, where they still largely overlap.}
    \label{Fig: compare_1}
\end{figure*}

\begin{figure*}
    \centering
    \includegraphics{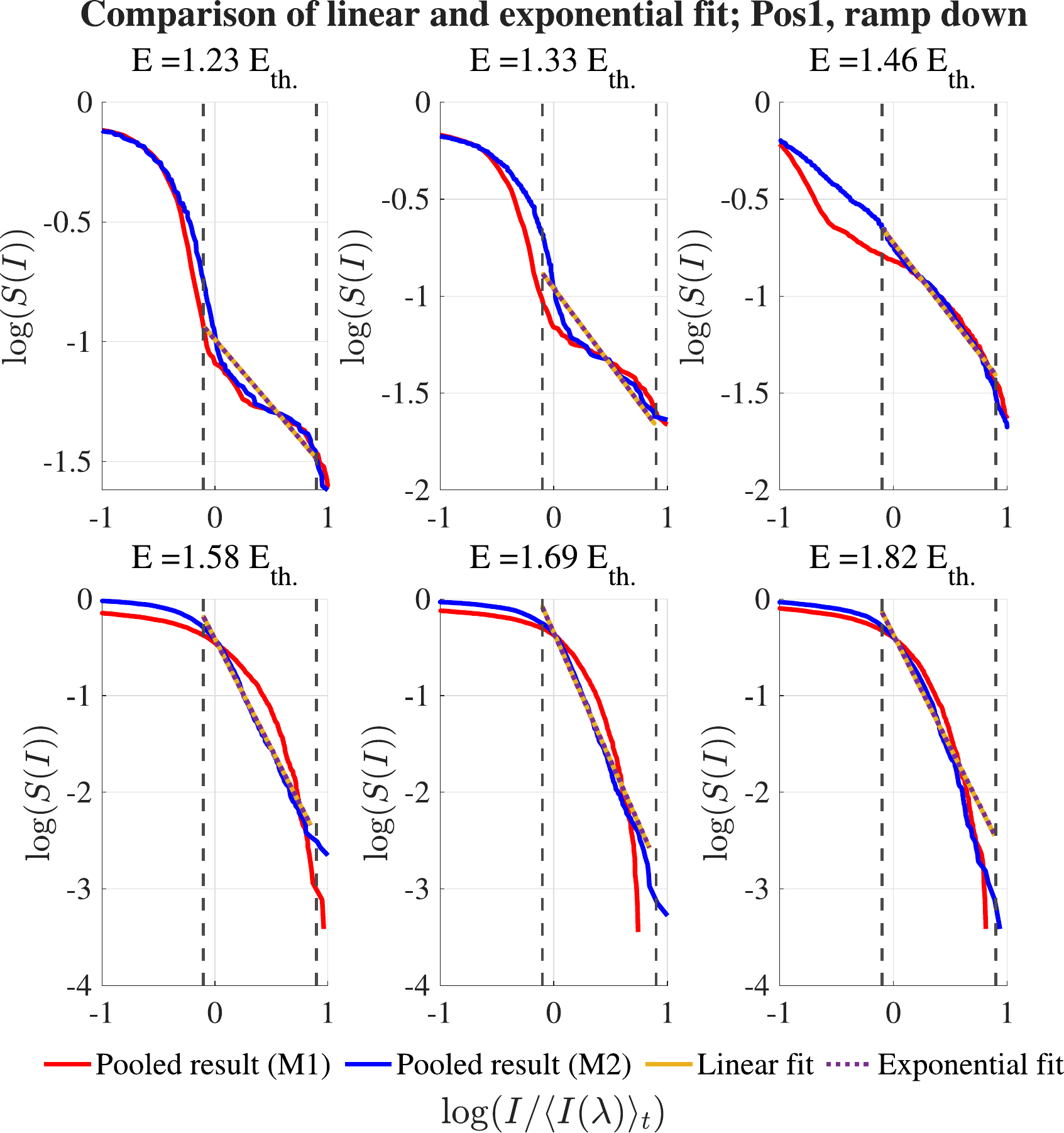}
    \caption{\textbf{Comparison of linear and exponential fit for Pos1, ramp down.} The fit is performed for M2. At the scale shown, linear fit (yellow) and exponential fit (purple) are identical.}    \label{Fig: compare_2}
\end{figure*}

\begin{figure*}
    \centering
    \includegraphics{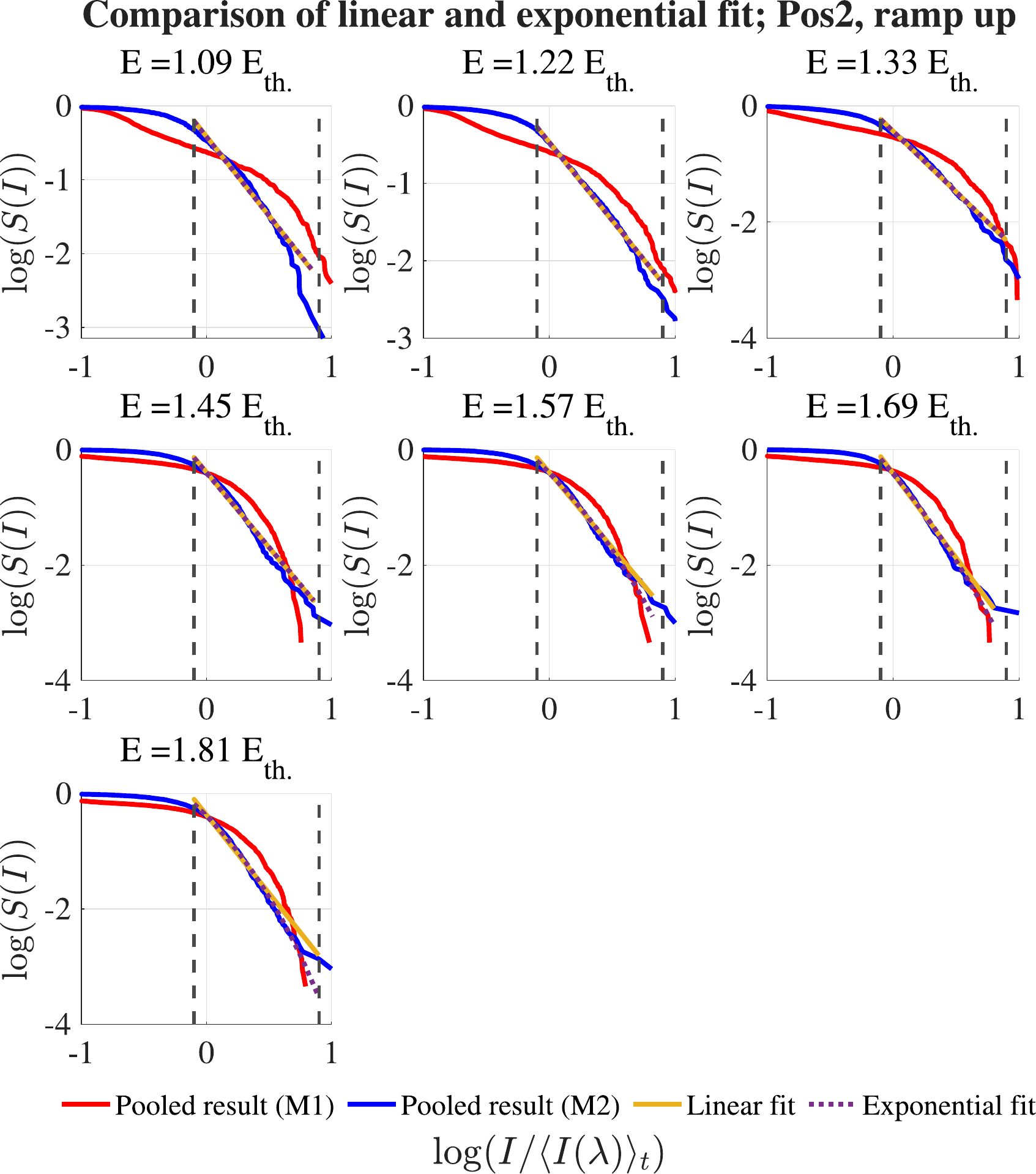}
    \caption{\textbf{Comparison of linear and exponential fit for Pos2, ramp up.} The fit is performed for M2. At the scale shown, linear fit (yellow) and exponential fit (purple) are the same for the majority of energies. For the last three energies ($E = 1.51 - 1.81E_{\text{th.}}$), there exists some small deviations. }
    \label{Fig: compare_3}
\end{figure*}

\begin{figure*}
    \centering
    \includegraphics{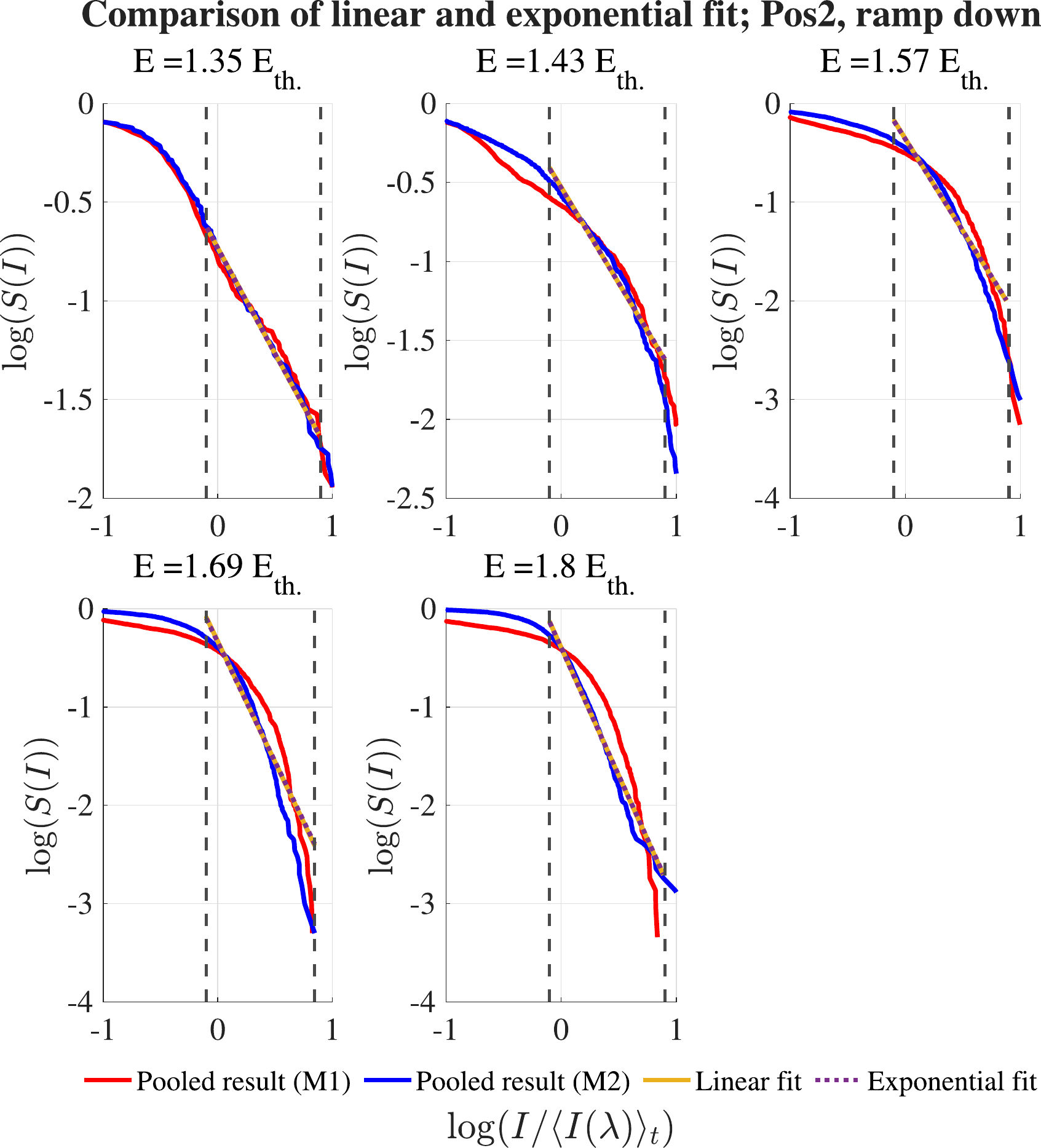}
    \caption{\textbf{Comparison of linear and exponential fit for Pos2, ramp down.} The fit is performed for M2. At the scale shown, linear fit (yellow) and exponential fit (purple) are identical.}    \label{Fig: compare_4}
\end{figure*}

We further show that an exponential decay representing the decay of a photonic mode in the weak coupling regime is \textit{not} a better description of the system in the region of interest. Fig.~\ref{Fig: compare_1} to~\ref{Fig: compare_4} show all the pooled survival functions that were generated and analyzed in the main text. In all cases (Pos1/Pos2, ramp up/down), linear and exponential fits to M2 are highly similar in log-log scale, meaning the region is appropriately described by a linear line. In conclusion, Lévy distribution is an adequate description of the random lasing system for high values of normalized intensities. 

\section{Evidence for the absence of photo-bleaching}

\begin{figure*}
    \centering
    \includegraphics{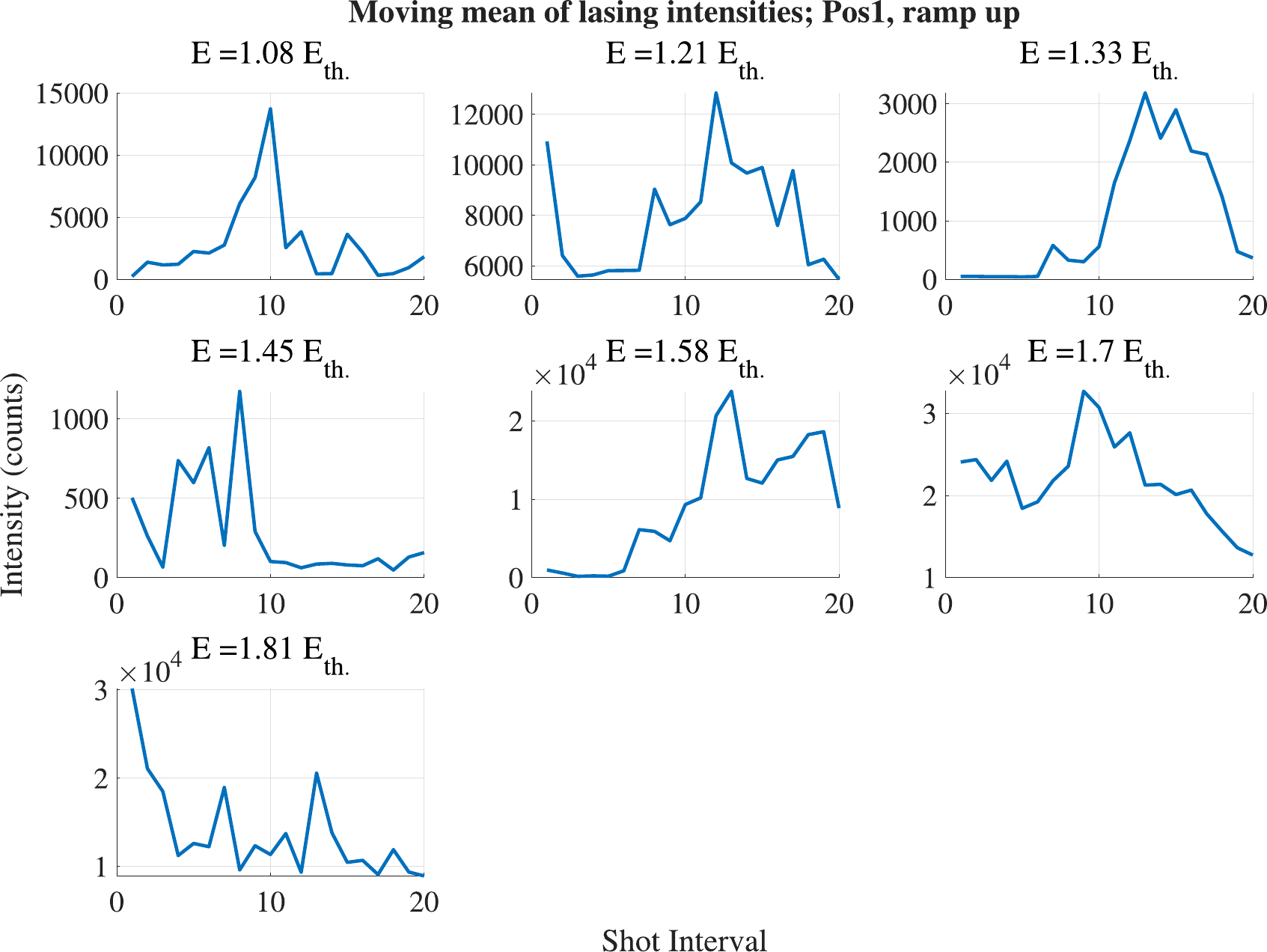}
    \caption{\textbf{Moving-window average of the total lasing intensities for Pos1, ramp up.} The titles for each panel are the pump energies. Only the data that is used for survival analysis is presented. Each shot interval consists of $10$ shots, translating to a $1$ second duration. All plots exhibit fluctuations during the acquisition time, demonstrating that there is no photo-bleaching of the sample.}
    \label{Fig: avg_tot_1}
\end{figure*}

\begin{figure*}
    \centering
    \includegraphics{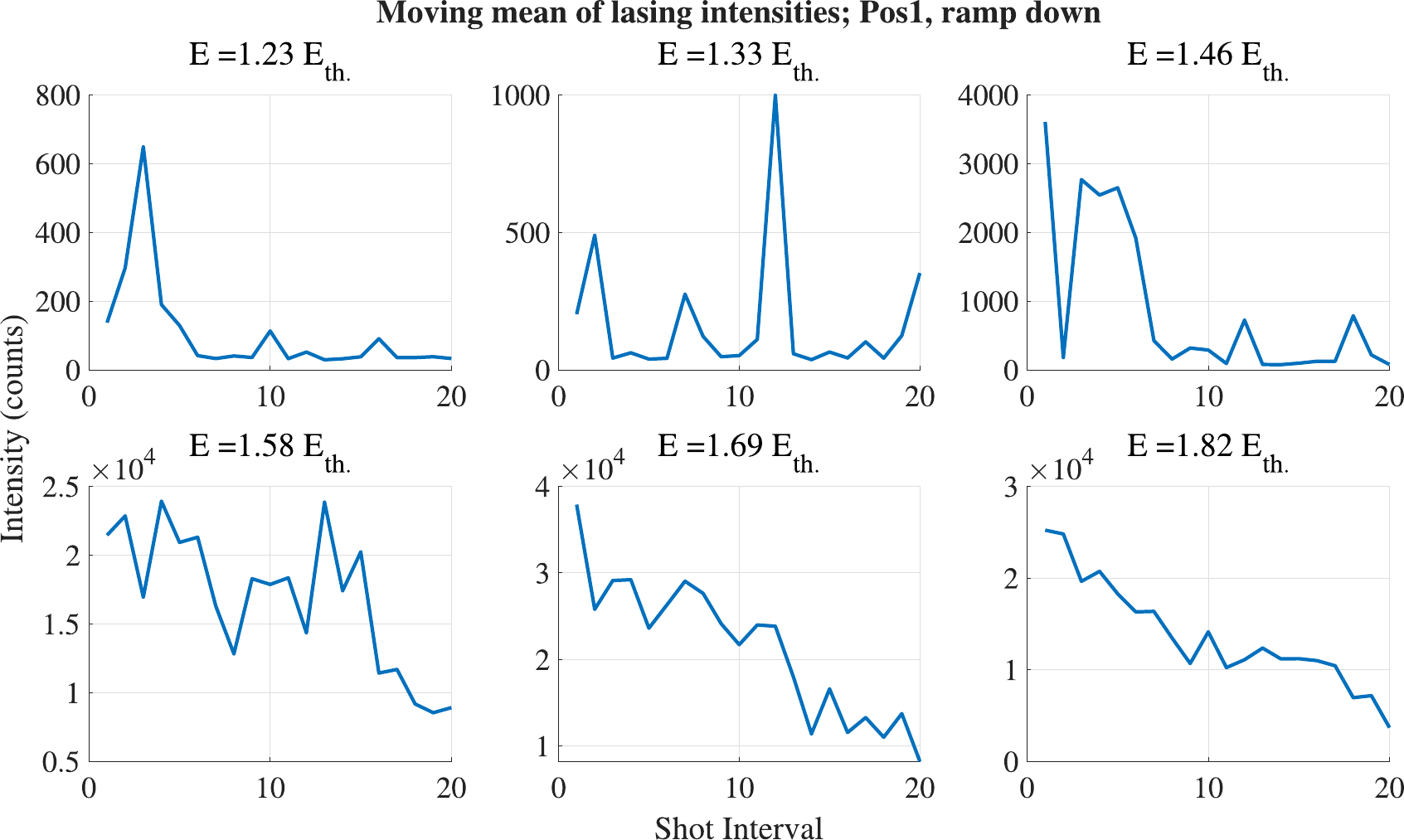}
    \caption{\textbf{Moving-window average of the total lasing intensities for Pos1, ramp down.} The titles for each panel are the pump energies. Only the data that is used for survival analysis is presented. Each shot interval consists of $10$ shots, translating to a $1$ second duration. All plots exhibit fluctuations during the acquisition time, demonstrating that there is no photo-bleaching of the sample.}
    \label{Fig: avg_tot_2}
\end{figure*}

\begin{figure*}
    \centering
    \includegraphics{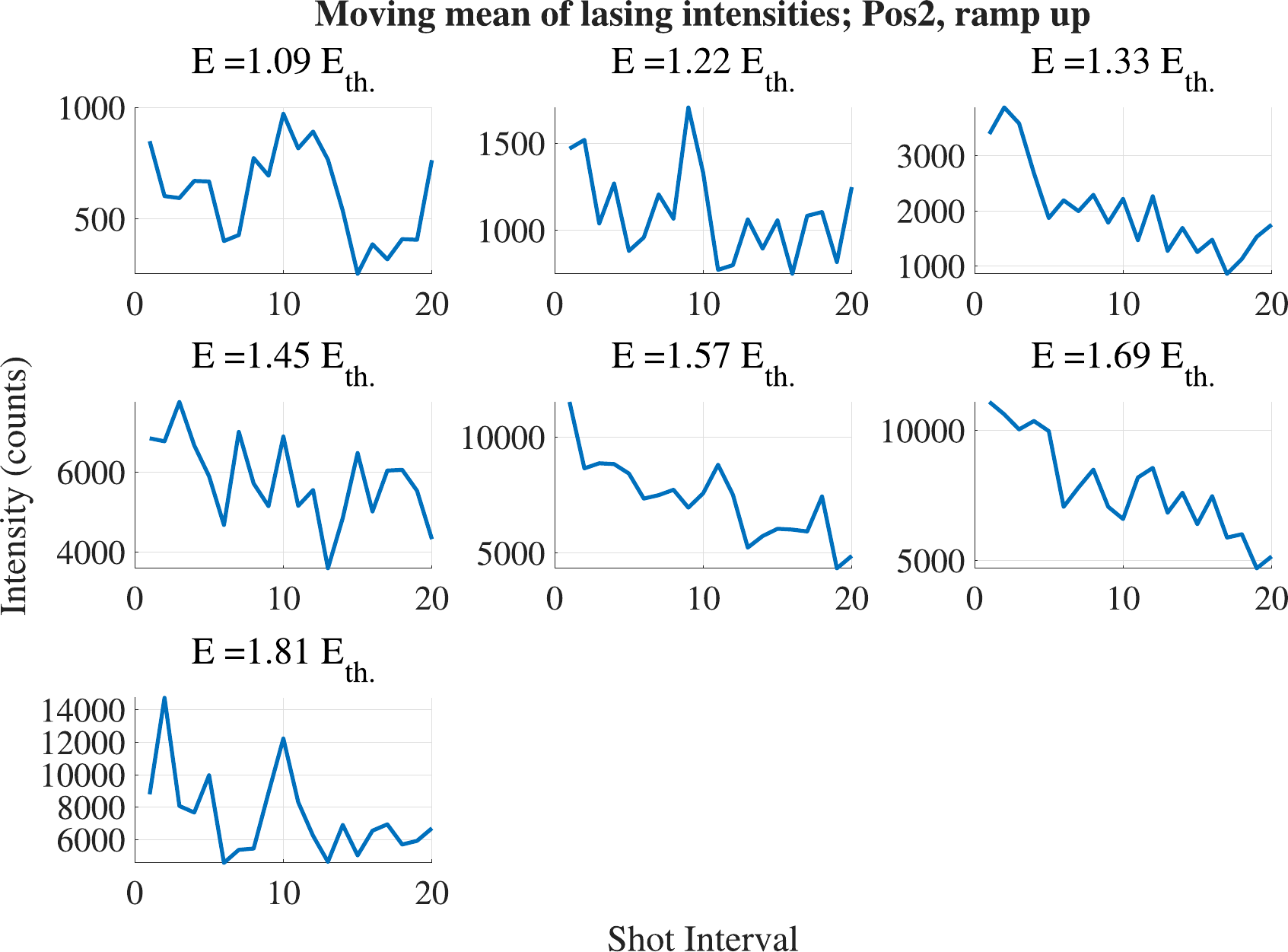}
    \caption{\textbf{Moving-window average of the total lasing intensities for Pos2, ramp up.} The titles for each panel are the pump energies. Only the data that is used for survival analysis is presented. Each shot interval consists of $10$ shots, translating to a $1$ second duration. All plots exhibit fluctuations during the acquisition time, demonstrating that there is no photo-bleaching of the sample.}
    \label{Fig: avg_tot_3}
\end{figure*}

\begin{figure*}
    \centering
    \includegraphics{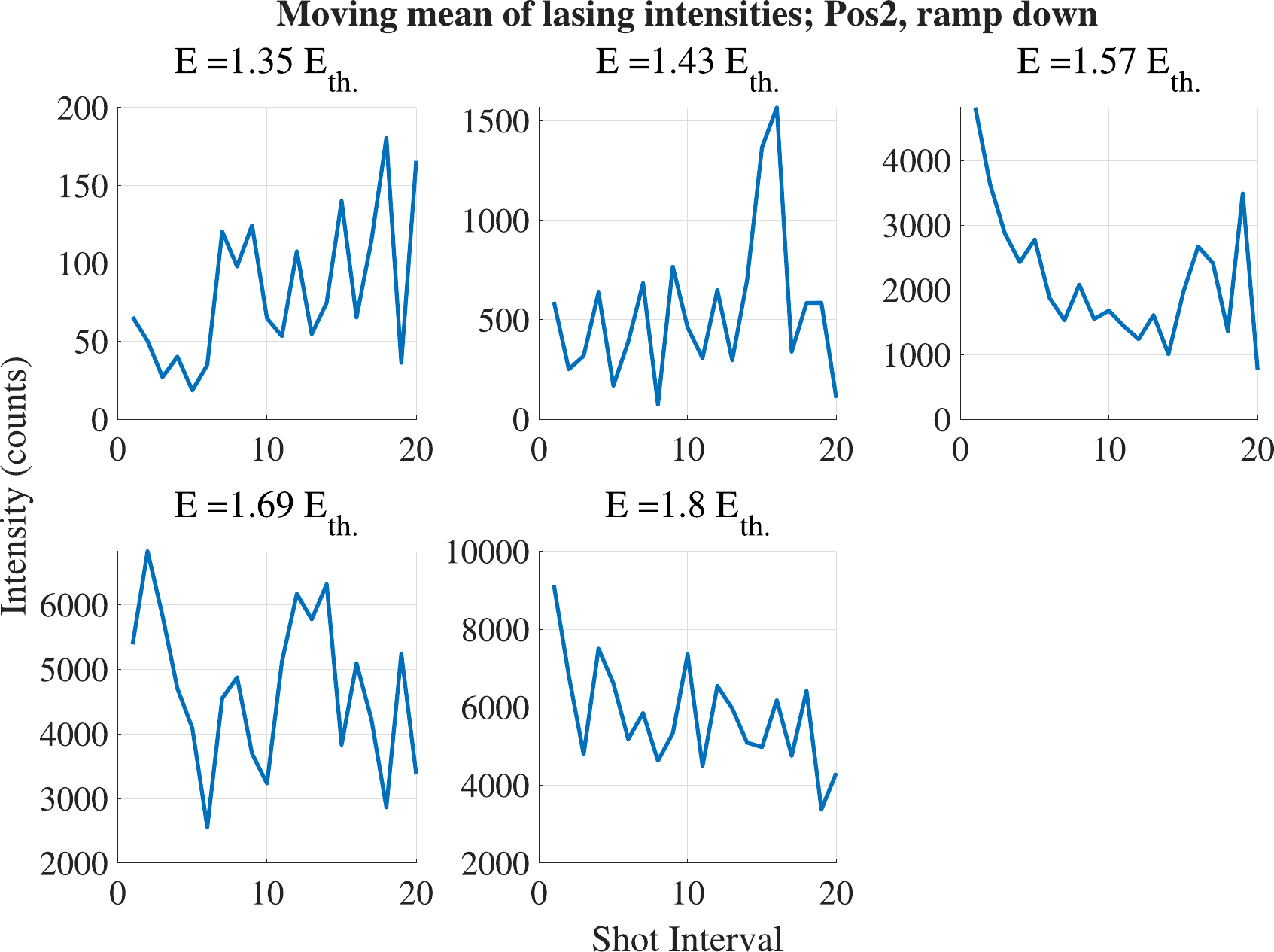}
    \caption{\textbf{Moving-window average of the total lasing intensities for Pos2, ramp down.} The titles for each panel are the pump energies. Only the data that is used for survival analysis is presented. Each shot interval consists of $10$ shots, translating to a $1$ second duration. All plots exhibit fluctuations during the acquisition time, demonstrating that there is no photo-bleaching of the sample.}
    \label{Fig: avg_tot_4}
\end{figure*}

Here we give evidence that the hysteresis observed in the main text is not due to photo-bleaching. One signature of photo-bleaching the laser dye is the monotonic decrease in lasing intensities over time due to light induced chemical changes in the dye molecules~\cite{demchenko_photobleaching_2020}. Fig.~\ref{Fig: avg_tot_1} to~Fig. \ref{Fig: avg_tot_4} present the moving-window average of the total lasing intensities for Pos1 and Pos2. The lasing modes are those selected for the survival analysis, and for each pump energy, which is marked by the titles on the panels, modes intensities are summed to produce the total intensities. Both the ramp up and down phases are included. The non-overlapping window size is $10$, resulting in $20$ shot intervals. In all of the data presented, the average total lasing intensities fluctuate randomly with time, and thus we conclude that there is no photo-bleaching of the RLs during the experiments. 

\section{Spectra evolution for Pos2}

To prove that the hysteresis that we have seen for Pos1 is reproducible, we present the qualitative single-shot spectra evolution for Pos2 in Fig.~\ref{Fig: evolution_2}. The lasing wavelengths for Pos2 are only minimally red shifted; from the spectra, we see two dominant modes taking the bulk of the gain in the system and becoming even more pronounced as we ramp up the incident energy (\textbf{a1} $\rightarrow$ \textbf{a7}). In the initial stage of the ramp down phase, the lasing persistence becomes diminished, as it can be seen from panels \textbf{b7} and \textbf{b6}. When the incident energy decreases beyond $1.46 E_{\text{th.}}$ (\textbf{b4}), lasing becomes entirely absent from that point on, where we only have occasional bursts during the entire $20$ seconds of acquisition window. At $1.1 E_{\text{th.}}$ (\textbf{b1}), the spectra become noise because the counts at all wavelengths do not exceed the noise level, affirmed by the ``fuzziness" of the colormap. At Pos2, we do not observe an outlier that embodies high correlations and persistence at a relatively low incident energy values, suggesting that the reasons that cause it in Pos1 is coincidental. 


\section{Possible physical origin of the synergy between correlation and persistence}

In this section, we discuss a potential microscopic mechanism by which correlations and persistence are coupled to each other based on theoretical works outlined in Ref.~\cite{tiwari_frequency_2012} and~\cite{uppu_identification_2012}. Although these references do not take into account LSPR that the AuNPs in our random lasers support, they illuminate key aspects of physics in this complex system.

\begin{figure*}
    \centering
    \includegraphics{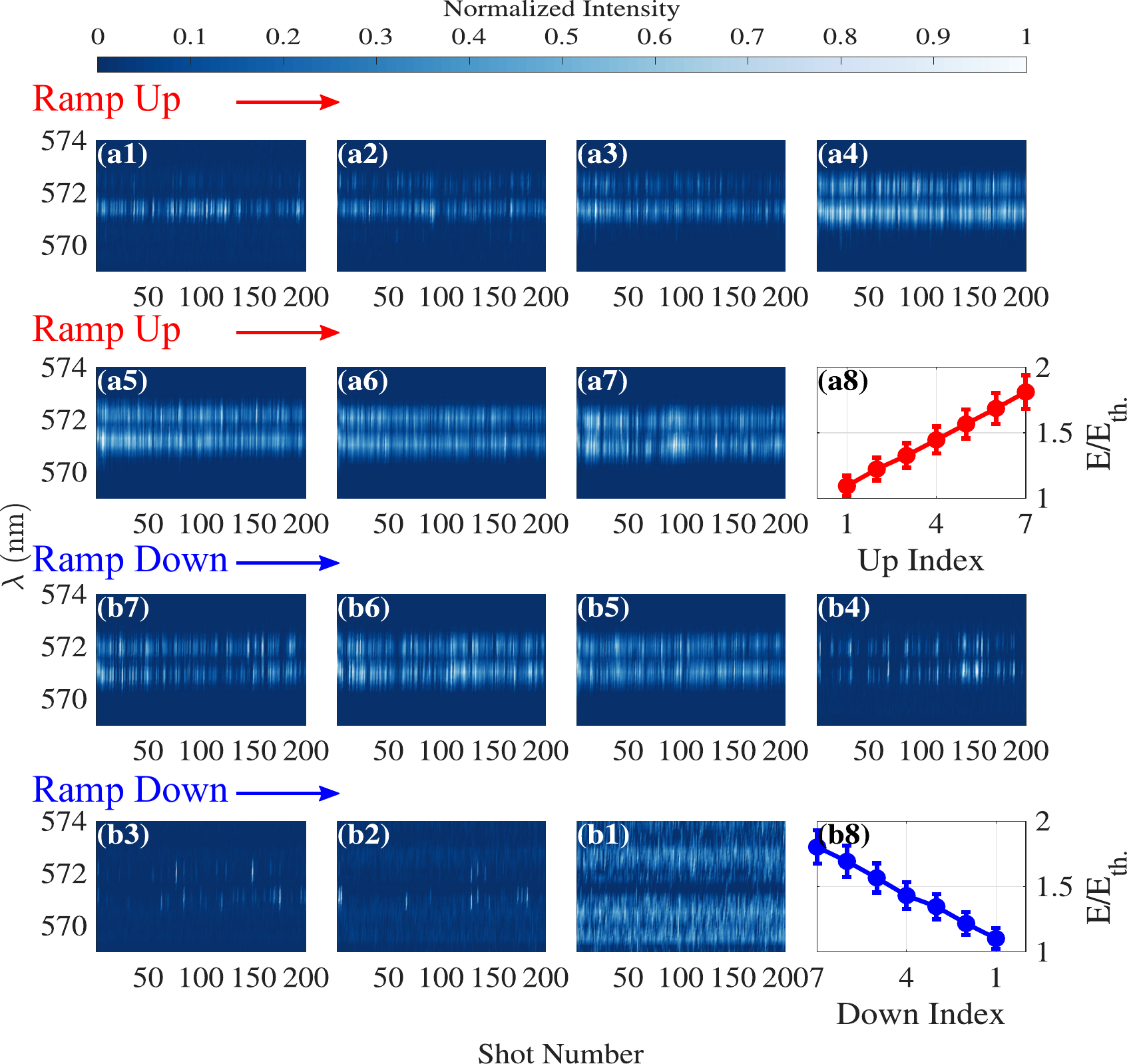}
    \caption{\textbf{The qualitative evolution of the RL for varying incident pump energy (Pos2).} \textbf{(a1) -- (a7),} Single-shot spectra in the ramp up phase. The energy range covered is $1.1 E_{\text{th.}} \rightarrow 1.8 E_{\text{th.}}$ in increment of $0.11 E_{\text{th.}}$. \textbf{(b7) -- (b1),} The corresponding spectra in the ramp down phase, starting at the highest energy. \textbf{(a8) and (b8),} Incident energy versus indices in the two phases, with the same index denoting the same energy values. The intensities are normalized with respect to each individual single-shot spectra. Note that \textbf{b1} panel has a fuzzy colormap because only noise is collected at this energy. Similar to Pos1, \textbf{a5} -- \textbf{a7} are similar to \textbf{b5} -- \textbf{b7} in terms of the lasing persistence and clear contrast, whereas we almost have no lasing from \textbf{b3} to \textbf{b1} in the ramp down phase. Therefore, Pos2 exhibits the same form of hysteresis as Pos1.}
    \label{Fig: evolution_2}
\end{figure*}

In the first article, Tiwari, Uppu, and Mujumdar numerically studied coherent random lasing in a diffusive system comprising monodisperse, resonant scatterers. Their Monte Carlo simulation reveals that the diffusive medium with intra-scatterer gain (Mie resonances) produce stabler lasing modes in the frequency domain, because there is a larger discrepancy between resonant and non-resonant photon dwell time ($\tau$), so only selected modes experience substantial, sustained gain. Our random laser supports intra-scatterer gain through the Mie resonances of LCs and AuNPs. As mentioned in the main text, the lasing wavelength interval remains around $571$ to $573$ nm and does not fluctuate appreciably over time (the same conclusion is reached for Pos2, see Fig. \ref{Fig: evolution_2}). Thus, it is hypothesized that high values of $\tau$ are concentrated on few lasing modes in our random laser, providing a physical explanation for the corresponding high extent of correlation. 

The second article by Uppu and Mujumdar explains the persistent lasing in coherent random lasers consisting monodisperse scatterers ($\text{TiO}_2$) with single-particle resonances through photon dynamics in disorder. Based on the numerical simulation, resonant photons (of the Mie resoances from the microspheres) dominate the contribution to the emission from resonant scatterers random laser across all pathlengths, even eclipsing light at the maximum gain cross section. Furthermore, there are  ``lucky photons'' with high pathlengths manifested as intense coherent peaks that do not drastically shift their wavelengths over time. Our random laser consists of similar components, exhibits wavelength selectivity, and have prominent modes that are long-standing. The fact that the amplified photons experience the gain from the same source --- Mie resonances --- is a plausible explanation for the coupling between correlation and persistence. 

Through reviewing theoretical works on the simplified version of our given system, we propose that the link between persistence and correlation could be attributed to emitted photons 1) having, on average, much higher $\tau$ than those that are suppressed and/or 2) experiencing the same kind of resonant feedback and dominating the output spectrum. Further simulations and experiments are needed to fully establish the role that LSPR plays in correlations and persistence.

\bibliography{reference}

\vfill
\pagebreak
